\documentclass[%
reprint,
 amsmath,amssymb,
 aps,
prb,
]{revtex4-2}

\usepackage{graphicx}% Include figure files
\usepackage{dcolumn}% Align table columns on decimal point
\usepackage{bm}% bold math
\usepackage{physics}
\usepackage{amsmath}
\usepackage{orcidlink}
\newcommand{\del}{\,{\mathit{\Delta}}}

\begin{document}

%\preprint{APS/123-QED}

\title{Fast and Accurate Prediction of Material Properties with Three-Body Tight-Binding Model for the Periodic Table}% Force line breaks with \\

\author{Kevin F. Garrity \orcidlink{0000-0003-0263-4157}}
\email{kevin.garrity@nist.gov}
 \author{Kamal Choudhary \orcidlink{0000-0001-9737-8074}}
 \affiliation{Materials Measurement Laboratory, National Institute of Standards and Technology, Gaithersburg MD, 20899}


\date{\today}% It is always \today, today,https://www.overleaf.com/project/6170b2c8c323b37ff64a893d

\begin{abstract}

 Parameterized tight-binding models fit to first principles calculations can provide an efficient and accurate quantum mechanical method for predicting properties of molecules and solids. However, well-tested parameter sets are generally only available for a limited number of atom combinations, making routine use of this method difficult. Furthermore, many previous models consider only simple two-body interactions, which limits accuracy. To tackle these challenges, we develop a density functional theory database of nearly one million materials, which we use to fit a universal set of tight-binding parameters for 65 elements and their binary combinations. We include both two-body and three-body effective interaction terms in our model, plus self-consistent charge transfer, enabling our model to work for metallic, covalent, and ionic bonds with the same parameter set. To ensure predictive power, we adopt a learning framework where we repeatedly test the model on new low energy crystal structures and then add them to the fitting dataset, iterating until predictions improve. We distribute the materials database and tools developed in this work publicly.
\end{abstract}

\maketitle

%\tableofcontents

\section{\label{sec:level1}Introduction}

With the growth in computing power over the past several decades, first principles electronic structure calculations have come to play an ever larger role in materials physics and materials design \cite{harrison2012electronic,martin2020electronic}. The increasing use of high-throughput computing techniques has allowed the construction of several databases containing calculated properties for thousands of materials \cite{curtarolo2013high,curtarolo2012aflow,jain2013commentary,kirklin2015open,choudhary2020joint,andersen2021optimade}. Nevertheless, there remain many types of calculations that are too computationally expensive to consider systematically, even at the level of relatively inexpensive semi-local density functional theory (DFT). Examples of these calculations include harmonic and anharmonic phonons \cite{wang1990tight}, thermal conductivity \cite{katre2016orthogonal}, thermoelectrics \cite{lee2006tight}, defect energetics \cite{kwon1994transferable} , surfaces \cite{mehl1996applications}, grain-boundaries \cite{morris1996tight}, phase-diagrams \cite{colinet1992tight,sluiter1988tight},  disordered materials \cite{roth1973tight}, dopants \cite{robertson1983dopant}, structure prediction \cite{chuang2006structure}, and molecular dynamics \cite{goedecker1995tight}. 

Building models based on DFT calculations is a major way to bridge the gap between existing databases and new properties or structures, but models are often developed on a case-by-case basis for single materials systems, which doesn't scale easily for materials design applications. Machine learning approaches \cite{vasudevan2019materials} with limited physics built-in have emerged in recent years as a very promising way to incorporate the large amount of DFT data available, but they can have difficulty extrapolating beyond their training data to new situations \cite{meredig2018can}. In this work, we aim to develop a physics-based model of the energy and electronic structure of materials, which we fit to a large database of DFT calculations using a combination of traditional and machine learning-inspired approaches.

Our underlying model is a tight-binding (TB) model where the TB Hamiltonian depends on a parameterized function of the crystal structure \cite{slater1954simplified,mehl1996applications,porezag1995construction,koskinen2009density,bannwarth2019gfn2,groth2014kwant,yusufaly2013tight,hourahine2020dftb+,seifert1996calculations,elstner1998self,kohler2005density,pu2004combining,vasudevan2019materials,schleder2019dft,schmidt2019recent,dftb_pt1, dftb_pt2, GFNxTB, vdw}, including the effects of charge self-consistency \cite{elstner2007scc,elstner1998self,frauenheim2000self}. This formalism contains the minimal description of quantum mechanics and electrostatics necessary to describe chemical bonding. The difficulty with this approach is producing a model that is both simple to fit and and efficient to evaluate while maintaining predictive accuracy.  Here, we go beyond previous works through a combination of two ideas. First, in addition to the typical two-body (two-center) atom-atom interactions, we use three-body (three-center) terms\cite{threecenter,fireball,PhysRevB_71_235101,PhysRevB_40_3979,PhysRevB_56_6594} to predict the tight-binding Hamiltonian from atomic positions. Including explicit three-body terms allows the Hamiltonian matrix elements between a pair of atoms to be environment dependent\cite{env1,env2,env3,env4}. This creates a more transferable model that can be applied with equal accuracy to many crystal structures and that better takes advantage of the abundance of DFT data available from modern computational resources. Previously, three center expansions have been used  most prominently\cite{PhysRevB_40_3979, PhysRevB_56_6594,  PhysRevB_71_235101} to approximate the exchange-correlation terms in tight-binding approaches that expand specific interactions from DFT\cite{elstner1998self, seifert2007tight, dftb,  porezag1995construction,fireball}. We instead include three-body interactions as general fitting parameters for both onsite and intersite matrix elements.

Second, we fit coefficients for 65 elemental systems (the main group and transition metals) as well as any binary combination of those elements, resulting in 2080 combinations. Within our framework, materials with three or more elements can be treated, but require three-body interactions between three different elements that go beyond the fitting in this work. Our total database consists of over 800,000 DFT calculations. Furthermore, we employ an active learning-inspired approach to continue generating new fitting data until our model performs well on out-of-sample tests. By fitting our model to a wide range of elemental and binary compounds, we hope to make a model that can be used in high-throughput or on-demand computing applications that are not possible with individually fit tight-binding models. Given a crystal structure, our three-body tight-binding model can calculate the band structure, total energy, forces, and stresses at a fraction of the computational cost of a direct DFT calculation. This combination of built-in physics, accuracy, and scope should allow our model to be applied for various materials design applications that are difficult with other techniques. 

We distribute a publicly available implementation of the present work and the fitting parameters at \url{https://github.com/usnistgov/ThreeBodyTB.jl} in the Julia programming language, as well as a python interface at \url{https://github.com/usnistgov/tb3py}. The documentation is available at \url{https://pages.nist.gov/ThreeBodyTB.jl/}, including examples. 
%The DFT database is available at XYZ website.% 

This work is organized as follows. Sec.~\ref{sec:tb} presents our tight-binding formalism, Sec.~\ref{sec:initial} describes a method to generate initial TB parameters for a single material via atomic projection, Sec.~\ref{sec:fit} provides details of the fitting process and dataset generation, Sec.~\ref{sec:results} shows tests of the model energy and electronic structure, and Sec.~\ref{sec:summary} presents conclusions.

\section{\label{sec:tb}Tight-binding formalism}
\subsection{Overview \label{sec:overview}}

The basic idea of TB is to perform electronic structure calculations in a minimal basis \cite{harrison2012electronic}. For example, a calculation of \textit{fcc} Al will have one $s$-orbital and three $p$-orbitals, for a total of four basis functions, rather than potentially hundreds of plane-waves or similar basis functions. Given a DFT calculation for a particular material, it is possible to use Wannier functions \cite{marzari2012maximally, mostofi2008wannier90, garrity2021database} or related techniques\cite{quambo, shirley, shirley2} to generate tight-binding Hamiltonians for that material. However, our goal is to predict the Hamiltonian directly from the crystal structure without performing an expensive DFT calculation first, allowing us to inexpensively predict the energy, band structure, and related properties.

Our tight-binding model is largely similar to formalism from density functional tight-binding including charge self-consistency \cite{elstner2007scc,elstner1998self,frauenheim2000self}, as well as the Navy Research Lab (NRL) tight-binding formalism \cite{mehl1996applications}. Here, we only include a brief overview of standard aspects of tight-binding, interested readers can consult the  review article such as \cite{goringe1997tight,frauenheim2002atomistic,koskinen2009density,seifert2007tight,elstner2007scc} for a more pedagogical introduction.

In addition to the band structure, we need to be able calculate the total energy, $E$. Many tight-binding formalisms make a distinction between the band structure and non-band structure contributions to the total energy, with the latter grouped together as a repulsive energy contribution, $E_{rep}$. We instead follow the NRL philosophy of grouping all the energy terms together by shifting the DFT eigenvalues, $\epsilon_i$,
\begin{eqnarray}
    E &=& \sum_i^{occ.} \epsilon_i + E_{rep} = \sum_{i}^{occ.} \epsilon'_{i} \\
    \epsilon'_i &=& \epsilon_i + E_{rep} / N
\end{eqnarray}
where $\epsilon'_i$ are the shifted eigenvalues and $N$ is the total number of electrons. After performing this shift, there is no need for a separate repulsive energy term. Below, we assume this shift has been done and do not write the prime explicitly.

We use non-orthogonal basis orbitals, where the tight-binding orbitals $\phi_\mu$ have a non-trivial overlap matrix $S_{\mu \nu} = \bra{\phi_{\mu}}\ket{\phi_{\nu}} $. The Hamiltonian is also a matrix $H_{\mu \nu} = \bra{\phi_\mu}  H \ket{\phi_\nu}$. The eigenvectors, $\psi_i = \sum_\mu c_\mu^i \phi_\mu$, with coefficients $c_\mu^i$ and eigenvalues, $\epsilon_i$, come from solving a generalized eigenvalue equation $H \psi_i = \epsilon_i S \psi_i$. The total energy is 
\begin{eqnarray}
    E &=& \sum_i f_i \sum_{\mu \nu} c_\mu^{i*} c_\nu^i H_{\mu \nu}
\end{eqnarray}
where $f_i$ is the occupancy of eigenstate $i$. For periodic systems, there is also an average over k-points, which is implicit above.

\begin{figure}
\includegraphics[width=3.4in]{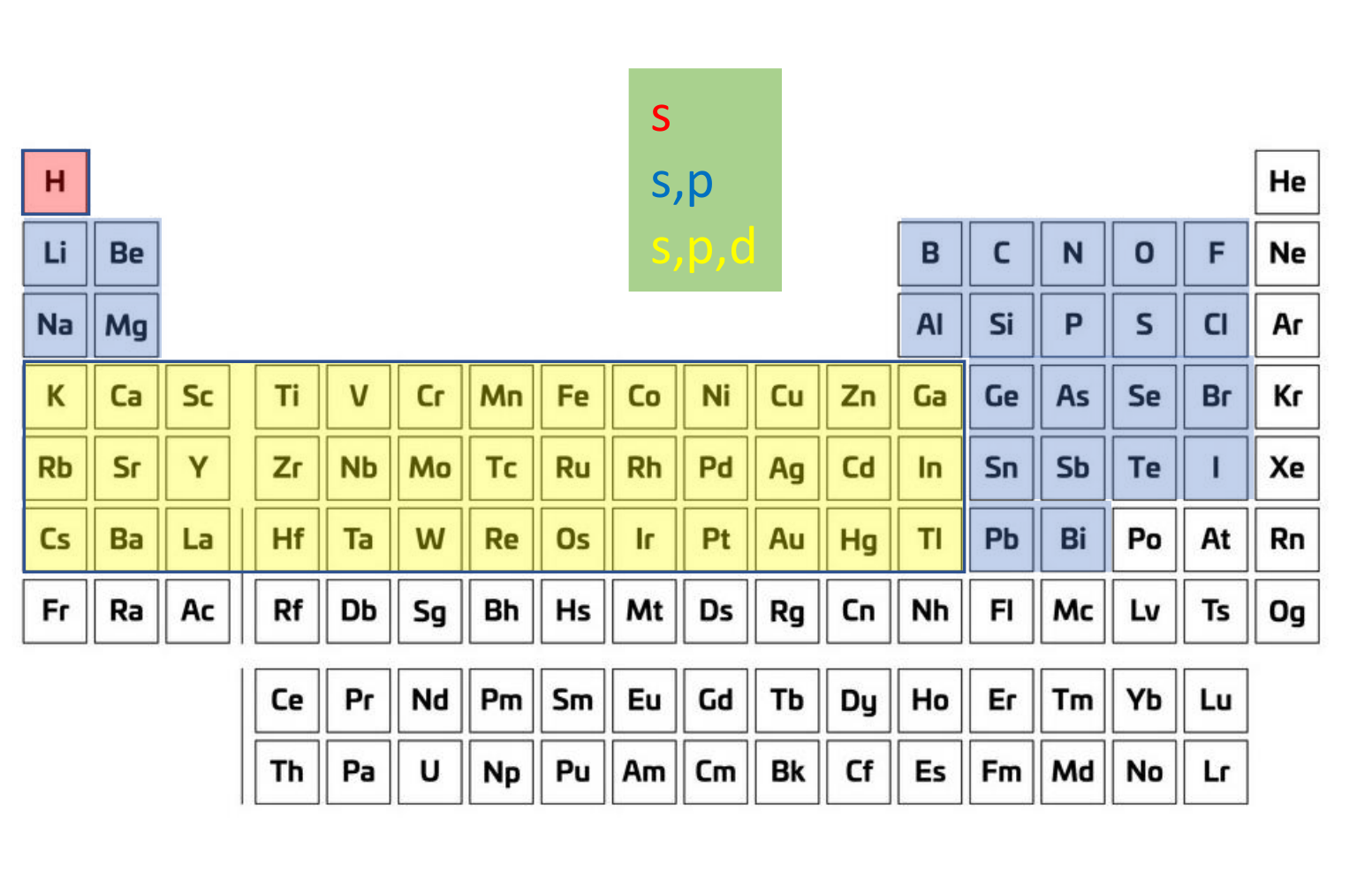}
\caption{\label{fig:s1} Orbitals used in our tight-binding model. Red: $s$ only (hydrogen), Blue: $sp$, Yellow: $spd$, White: Not included in model.}
\end{figure}

Once we have the Hamiltonian, solving the model involves diagonalizing a matrix with four ($sp$) or nine ($spd$) basis functions per atom, which is computationally inexpensive for small-to-medium sized systems. The orbitals we chose for each element are listed in Fig.~\ref{fig:s1}. The overlap matrix can be fit easily from the atomic orbitals. Thus, predicting a set of matrix elements, $H_{\mu \nu}$, that accurately reproduce the energy and band structure directly from the crystal structure is the main challenge of developing a parameterized tight-binding model.

\subsection{Charge self-consistency\label{sec:scc}}

A major limitation of the above formalism is that it does not include any explicit role for charge transfer or the resulting long-range Coulomb interaction. While this may be adequate for elemental systems and some metal alloys, explicitly including self-consistent electrostatics greatly improves fitting for ionic systems, as the remaining interactions become short-ranged\cite{koskinen2009density,elstner2007scc,elstner1998self,frauenheim2000self}. In this work, we do not consider magnetism, but spin self-consistency can be included along similar lines. The cost of including self-consistency is that the eigenvalue problem will have to be solved several times to reach convergence, in a manner similar to solving the Kohn-Sham equations. In practice, the smaller basis sets used in tight-binding reduce the convergence difficulties, and similar charge mixing schemes can be employed\cite{pulay}.

The key variable for charge self-consistency is $\del q_I$, the excess charge on ion $I$, relative to a neutral atom:
\begin{eqnarray}
    q_I &=& \sum_i f_i \sum_{\mu \in I} \sum_\nu \frac{1}{2} (c_\mu^{i*} c_\nu^i + c_\mu^{i} c_\nu^{i*}) S_{\mu \nu} \\
    \del q_I &=& q_I - q_I^0
\end{eqnarray}
where $q_I^0$ is the valence ionic charge. $\del q_I$ enters the expression for the Coulomb energy, $E_{coul}$,
\begin{eqnarray}
E_{coul} = \frac{1}{2} \sum_{IJ} \gamma_{IJ} \del q_I \del q_J
\end{eqnarray}
where $\gamma_{IJ}$ is the Coulomb operator:
\begin{eqnarray}
\gamma_{IJ} &=& 
\begin{cases} 
U_I & I = J \\
\frac{erf(C_{IJ} R_{IJ})}{R_{IJ}} & I \neq J
\end{cases}\\
C_{IJ} &=& \sqrt{\frac{\pi / 2}{1/U_I^2 + 1/U_J^2}}.
\end{eqnarray}
At long distances, $\gamma_{IJ}$ follows $1/R_{IJ}$, where $R_{IJ}$ is the distance between ions $I$ and $J$. $U_I$ is an onsite Hubbard term, which we fit to the changes in atomic eigenvalues for different numbers of electrons. The $erf(C_{IJ} R_{IJ})$ term reduces the interaction between nearby orbitals due to orbital overlap, and goes to 1 at long distances, please see details in \cite{elstner2007scc,elstner1998self,frauenheim2000self}.
 
Incorporating the Coulomb term, our expression for the total energy is now 
\begin{eqnarray}\label{eq:toten}
    E &=& \sum_i f_i \sum_{\mu \nu} c_\mu^{i*} c_\nu^i H_{\mu \nu} + \frac{1}{2} \sum_{IJ} \gamma_{IJ} \del q_I \del q_J
\end{eqnarray}
and the Hamiltonian used to calculate the eigenvectors and eigenvalues must be modified to 
\begin{eqnarray}
    H_{\mu \nu}' = H_{\mu \nu} + \frac{1}{2} S_{\mu \nu} \sum_K (\gamma_{IK} + \gamma_{JK}) \del q_K. 
\end{eqnarray}

\subsection{\label{sec:twobody}Two-body Intersite Interactions}

The largest contributions to the intersite Hamiltonian matrix elements $H_{\mu \nu}$ are the two-body interactions between orbitals $\mu$ and $\nu$. Following the Slater-Koster\cite{slater1954simplified} formalism, these terms can be factored into functions that depend solely on distance between the two atoms and symmetry factors that depend on the orbital types ($s$, $p$, or $d$) and their relative orientations. The symmetry factors are tabulated by the Slater-Koster matrix elements $M^x_{ij}$, where $i$ and $j$ are the orbitals, and $x$ is an index over a number of components (traditionally labeled $\sigma$, $\pi$, $\delta$):
\begin{eqnarray}\label{eq:twobody1}
H^{2bdy}_{iI, jJ} = \sum_x f^x_{iI,jJ}(R_{IJ}) M^x_{ij}. 
\end{eqnarray}
Here $H^{2bdy}_{iI, jJ}$ are the two-body Hamiltonian matrix elements between orbital $i$ on atom $I$ and orbital $j$ on atom $J$. These depend on $f^x(R_{IJ})$, which are functions of the distance between the atoms. We expand the function of distance in terms of the Laguerre polynomials $L_x(d)$ times a decaying exponential:
\begin{eqnarray}\label{eq:lag}
f_{iI,jJ}(d) = e^{-a d}\sum_{x} f_{iI,jJ}^x L_x(d), 
\end{eqnarray}
where $f_{iI,jJ}^x$ are fitting coefficients that depend on the types of atoms $I$, $J$ and the orbital types $i$, $j$. $a$ is a universal decay constant that is set to 2 Bohr $\approx$ 1.058 \AA. The Laguerre polynomials are chosen because they are complete and orthogonal with respect to the inner product $<\,f,g\,> = \int_0^\infty f(x) g(x) e^{-x} dx$ and result in numerically stable fits. We use five terms in the above expansion to fit the two-body Hamiltonian matrix elements. 

We use an identical formalism to fit the overlap matrix elements, except we use seven terms as there is less danger of overfitting. Unlike the Hamiltonian, the overlap matrix elements are approximating overlap integrals that are explicitly two-body,  so there is no need for three-body interactions.

The decay constant parameter $a$ can be optimized to improve the convergence speed of the Laguerre expansion. Because the overlaps themselves and the inter-site Hamiltonian are due to orbital overlap, the optimal choice for $a$ is close to the decay length of the valence atomic orbitals we include. These decay lengths are set by the valence orbital eigenvalues, and are therefore in same range for all elements. We find that any value near 1 \AA\,is reasonable and gives similar results. We note that by fixing the decay constant, the two-body Hamiltonian now depends linearly on the fitting coefficients $f_{iI,jJ}^x$, which greatly simplifies the fitting procedure. We will design the other terms in our model such that they are linear as well. 

\subsection{\label{sec:threebody}Three-body Intersite Interactions}

\begin{figure}
\includegraphics[width=3.4in]{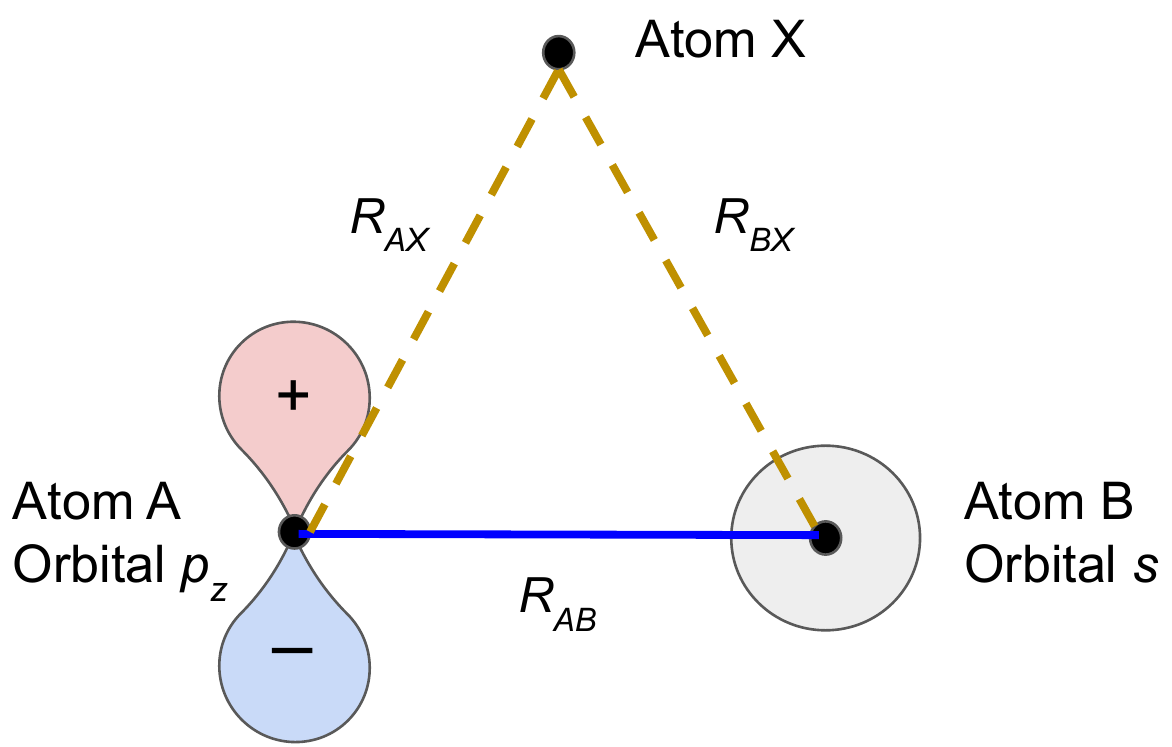}
\caption{\label{fig:schematic} Schematic of three-body terms. The direct two-body interaction between the $p_z$-orbital on atom A (left) and the $s$-orbital on atom B (right), represented by the solid blue line, is zero by symmetry. However, atom X (top) breaks the mirror symmetry and allows a non-zero $H_{p_z A, s  B}$ via the three body interaction (dashed lines).}
\end{figure}

Most tight-binding formalisms ignore contributions to the intersite Hamiltonian matrix elements that go beyond the two-body terms that we consider above. While this is usually adequate for fitting to a single structure at various volumes or with small distortions, it leads to well-known difficulties when fitting to multiple structures that we discus further in Sec.~\ref{sec:simple}. In such situations, the best matrix elements for each structure cannot be fit with a single function of distance. 

While there are various methods to alleviate this problem by including neighborhood-dependent hoppings\cite{env1,env2,env3,env4}, here, we directly include three-body terms in our fitting\cite{threecenter}. For example, consider $H_{p_z A, s  B}$, the interaction between the $p_z$-orbital on atom A and the $s$-orbital on atom B in Fig.~\ref{fig:schematic}. Due to the symmetry of the orbitals, the direct two-body interaction (solid line) is zero. However, the presence of other atoms, in this case atom X, will modify this interaction. Here, atom X allows a non-zero interaction by breaking the mirror symmetry along the line from A to B. This three-body interaction can be represented as two hoppings (dashed lines in Fig.~\ref{fig:schematic}): A to X, and then X to B. If we assume atom X has the symmetry of an $s$-orbital, then this pair of hoppings is indeed non-zero. Thus, including three-body interactions in this way allows atom X to modify the A-B interaction.

We implement this idea in our model by including a contribution to the intersite matrix elements from nearby third atoms :
\begin{eqnarray} 
H^{3bdy}_{iI, jJ} =  \sum_K g_{iI,jJ,K}(R_I, R_J, R_K) M_{is} M_{js}.\label{eq:threebody1} 
\end{eqnarray}
Here the sum over $K$ is a sum over nearby third atoms, and the symmetry factors are a product of two Slater-Koster symmetry factors, with the symmetry of the third atom assumed to be an $s$-orbital, \textit{i.e.} isotropic. This symmetry assumption can be viewed either as the simplest assumption or as the first term in an expansion, and it will not break any symmetries required by the space group. However, as discussed above, the three-body term can correctly split certain degeneracies or allow for non-zero couplings if those ``extra'' symmetries are artifacts of assuming a purely two-body interaction.

The fitting function $g$ can in principle depend on a complicated function of the three atom positions, which creates potential problems with over-fitting. In order to make progress, we make the simplifying assumption that the three-body terms can be expanded in terms of the three distances $R_{IJ}$, $R_{IK},$ and $R_{JK}$ only, and furthermore, only a few terms are necessary in the expansion:
\begin{equation}\label{eq:threebody2}
\begin{split}
&g_{iI,jJ,K}(R_I, R_J, R_K) = e^{-a(R_{IK} + R_{JK})} [ \\
&g_1 L_0(R_{IK}) L_0(R_{JK}) \\ + &g_2 L_0(R_{IK}) L_1(R_{JK}) \\ + &g_3 L_1(R_{IK}) L_0(R_{JK}) \\
+&g_4 L_0(R_{IK}) L_0(R_{JK}) L_0(R_{IJ}) e^{-a R_{IJ}}].
\end{split}
\end{equation}
Here, there are four fitting coefficients $g_i$ multiplied by specific products of Laguerre polynomials times decaying exponentials. The $g_i$ depend on the types of atom $I$, $J$, and $K$, as well as the orbitals $i$ and $j$, but we suppress these indexes for clarity. We find through experimentation that these four terms have the largest contribution in typical cases. In the case where atom $I$ and $J$ are the same type, there are only three independent coefficients, as $g_2=g_3$ by permutation symmetry. Importantly, the contribution from the third atom decays exponentially as it moves further away from either of the primary two atoms, which constrains the contributions to be short-ranged.

We note that the self-consistent electrostatic terms introduced in Sec.~\ref{sec:scc} can also create effective three-body interactions. However, there is no issue with double counting as the three-body body terms introduced in this section are fit after the effects of charge self-consistency have already been included.

\subsection{\label{sec:onsite}Onsite Interactions}

The onsite matrix elements $H_{iI, jI}$ require significant care to fit, as they effectively incorporate the contributions from the normal repulsive energy term (see section \ref{sec:overview}). The one-body term is due to the non-spin-polarized spherically symmetric atomic eigenvalues $\epsilon_{iI}$. The two-body terms modify the orbital energies due to a single nearby atom. They are split into an average term and a crystal-field term. The former changes the average eigenvalue of a set of orbitals (\textit{e.g.} $p$-orbitals) due to a nearby atom, while the latter can split the degeneracy of a set orbitals depending on the site symmetry. Finally, we include a simple three-body term discussed below:
\begin{eqnarray}
H_{iI, jI} &=& \epsilon_{iI} \delta_{ij} + H^{avg}_{iI} \delta_{ij} + H^{cf}_{iI, jI} + H^{3bdy}_{I} \delta_{ij}\\
H^{avg}_{iI} &=& \sum_J h_{iIJ}(R_{IJ}) \\
H^{cf}_{iI, jI} &=& \sum_J h_{iI,jJ}^{cf}(R_{IJ}) M_{is} M_{js} \\
\end{eqnarray}
Here, $\delta_{ij}$ is the Kronecker delta function, $H^{avg}_{iI, iI}$ is the average interaction, $H^{cf}_{iI, jI}$ is the crystal field interaction, and $H^{3bdy}_{iI, iI}$ is the three-body interaction. Like the two-body inter-atomic term (see Eq. \ref{eq:lag}), the average interaction is expanded as a Lagauerre polynomial times a decaying exponential. The crystal field term is very similar except it includes a pair of symmetry factors. Similar to the three-body intersite case discussed above, we assume the second atom contributes with isotropic $s$-orbital symmetry. The crystal field term allows the mixing of different orbitals on the same atom (\textit{e.g.} $s$ and $p_x$) if the atom is on a low symmetry site.
\begin{eqnarray}\label{eq:onsite2}
h_{iIJ}(d) &=& e^{-a d}\sum_{x} h_{iIJ}^x L_x(d) \\
h^{cf}_{iI,jJ}(d) &=& e^{-a d} \sum_x h_{iI,jJ}^{cf, x} L_x(d).
\end{eqnarray}
$h_{iIJ}^x$ and $h_{iI,jJ}^{cf,x}$ are the fitting coefficients for the average and crystal field terms, respectively. We fit them with four terms in the expansion ($x=1-4$).

Finally, there is a three-body average onsite interaction. To simplify the fitting, we apply this term to all orbitals on an atom equally, without an orbital dependence.
\begin{equation}
H^{3bdy}_{I} = \sum_{JK} h^{3bdy}_{IJK}(R_{IJ},R_{IK},R_{JK})  
\end{equation}
This is again expanded into four terms:
\begin{equation}\label{eq:onsite3}
\begin{split}
&h^{3bdy}_{IJK}(R_{IJ},R_{JK},R_{IK}) = e^{-a(R_{IJ}+R_{IK}+R_{JK})} \times [\\
&h^{3bdy}_1 L_0(R_{IJ}) L_0(R_{JK}) L_0(R_{IK}) + \\
&h^{3bdy}_2 L_1(R_{IJ}) L_0(R_{JK}) L_0(R_{IK}) + \\
&h^{3bdy}_3 L_0(R_{IJ}) L_1(R_{JK}) L_0(R_{IK}) + \\
&h^{3bdy}_4 L_0(R_{IJ}) L_0(R_{JK}) L_1(R_{IK}) ]. 
\end{split}
\end{equation}
$h^{3bdy}_x$ are the four fitting coefficients, which depend also on $IJK$. In the case where the type of atom $J$ and $K$ are the same, only three fitting coefficients are independent due to permutation symmetry. We discuss the relative magnitudes of typical three-body terms in an example material in supplementary materials Sec.~S3\footnote{See Supplemental Material at [URL will be inserted by
  publisher] for data tables, code timings, and additional examples}.

\section{\label{sec:initial}Atomic projection of wavefunctions}

\subsection{Projection Method\label{sec:atomproj}}

In order to fit the model defined in section \ref{sec:tb}, we need data from DFT calculations. While we will primarily concentrate
on fitting to energies and eigenvalues as discussed later, we need a reasonable set of tight-binding parameters to start the fitting process. 
A difficultly comes from the fact that even a set of isolated bands can be described by many different tight-binding models, as it is always possible to apply unitary transformations to a Hamiltonian without changing the eigenvalues. Furthermore, the conduction bands we wish to describe with tight-binding are generically entangled with both higher energy atomic levels and free-electron bands that we cannot describe with our model. Maximally-localized Wannier functions and similar methods \cite{marzari2012maximally, mostofi2008wannier90,quambo} are a well-known ways to generate a tight-binding Hamiltonian. However, because they require an optimization step, they are not guaranteed to resemble atomic-like orbitals in general cases, and they can depend discontinuously on atomic positions, making them a poor choice for the fitting data we need. Symmetry-adapted Wannier functions can improve the situation for some structures, but the same issues remain for broken symmetry structures\cite{symwan}.

We want a procedure to generate the best tight-binding matrix for our goal, which is to serve as the data for fitting the model described in Sec.~\ref{sec:tb}. We therefore use a non-iterative atomic orbital projection procedure. Projection schemes have the advantage of maintaining the correct symmetry of the tight-binding
Hamiltonian and do not require optimization. Following similar schemes\cite{PhysRevB.88.165127, PhysRevB.93.035104}, the basic idea involves projecting the large $N$-band Kohn-Sham Hamiltonian $H^{KS}$ at a given $k$-point onto a smaller number of $M$ atomic orbitals:
\begin{eqnarray}
H_{\alpha, \beta}^{TB} &=& \bra{\phi_{\alpha}}  H^{KS}   \ket{\phi_{\beta} }  \label{eq:proj1}\\
&\approx& \sum_n \bra{\phi_{\alpha}} \ket{\psi_n} E_n \bra{\psi_n}  \ket{\phi_{\beta} }, \label{eq:proj2} 
\end{eqnarray}
where $\phi_{\alpha}$ are atomic-like orbitals, and $\psi_n$ and $E_n$ are the Kohn-Sham eigenvectors and eigenvalues in a plane-wave basis.

A difficulty with Eq. \ref{eq:proj2} is how to select the best $M$ bands that are appropriate to describe with atomic-like orbitals in the case of entanglement, which is generic for conduction bands. We proceed by defining a set of projection coefficients $B_{\alpha, n} = \bra{\phi_\alpha}\ket{\psi_n}$. Then, we consider the projection matrix for eigenvectors:
\begin{eqnarray}
(B^{\dag} B)_{n,m} = \bra{\psi_n} P \ket{\psi_m} = P_{n,m}
\end{eqnarray}
The diagonal elements of this $N \times N$ matrix are the projectibility of each band\cite{PhysRevB.88.165127, PhysRevB.93.035104}. 

Our key approximation is to represent the projection matrix $P$ with a new matrix $\tilde{P}$, created from the $M$ eigenvectors of $P$ that have the largest eigenvalues. 
\begin{eqnarray}
P_{i,j} &=& \sum_{m,n=1}^N Q_{i,n} p_{n,m} (Q_{j,m})^\dag \\
\tilde{P}_{i,j} &=& \sum_{n=1}^M Q_{i,n} (Q_{j,n})^\dag \label{eq:pbar} \\
\tilde{P} &=& \tilde{B}^\dag \tilde{B} \label{eq:bbar}
\end{eqnarray}
Here, $Q_{i,n}$ are the $N$ eigenvectors of $P$, and $p_{n,m}$ are a diagonal matrix of eigenvalues.
The sum in Eq.~\ref{eq:pbar} is over the $M$ largest eigenvalues, and Eq.~\ref{eq:bbar} defines $\tilde{B}$, which is a $M \times N$ matrix. $\tilde{P}$ projects onto the highest projectiblity $M$-dimensional subspace to represent the $M$ atomic wavefunctions. By construction, it has $M$ eigenvalues equal to 1, with the rest equal to zero. Using $\tilde{P}$, we can now apply the philosophy of Eq.~\ref{eq:proj2} without difficulty:   
\begin{eqnarray}
H^{TB} &=&   B \tilde{P} E \tilde{P}^\dag B^\dag \\
H^{TB} &=&  B \tilde{B}^\dag \tilde{B} E \tilde{B}^\dag \tilde{B} B^\dag
\end{eqnarray}
Here, $E$ is a diagonal $N \times N$ matrix of the original eigenvalues. 

By approximating $P$ with its $M$ eigenvectors with large eigenvalues, we have effectively selected the $M$-dimensional subspace of the original larger Hamiltonian that are best (most atomic-like), thus avoiding the difficulty of the naive Eq.~\ref{eq:proj2}. This projection scheme can then be applied to a grid of $k$-points, and the resulting TB Hamiltonian can be Fourier-transformed onto a real-space grid. Because the original atomic-like states are localized in real-space, the real-space Hamiltonian will also be localized as well, although not maximally-localized.

\begin{figure}
\includegraphics[width=3.0in]{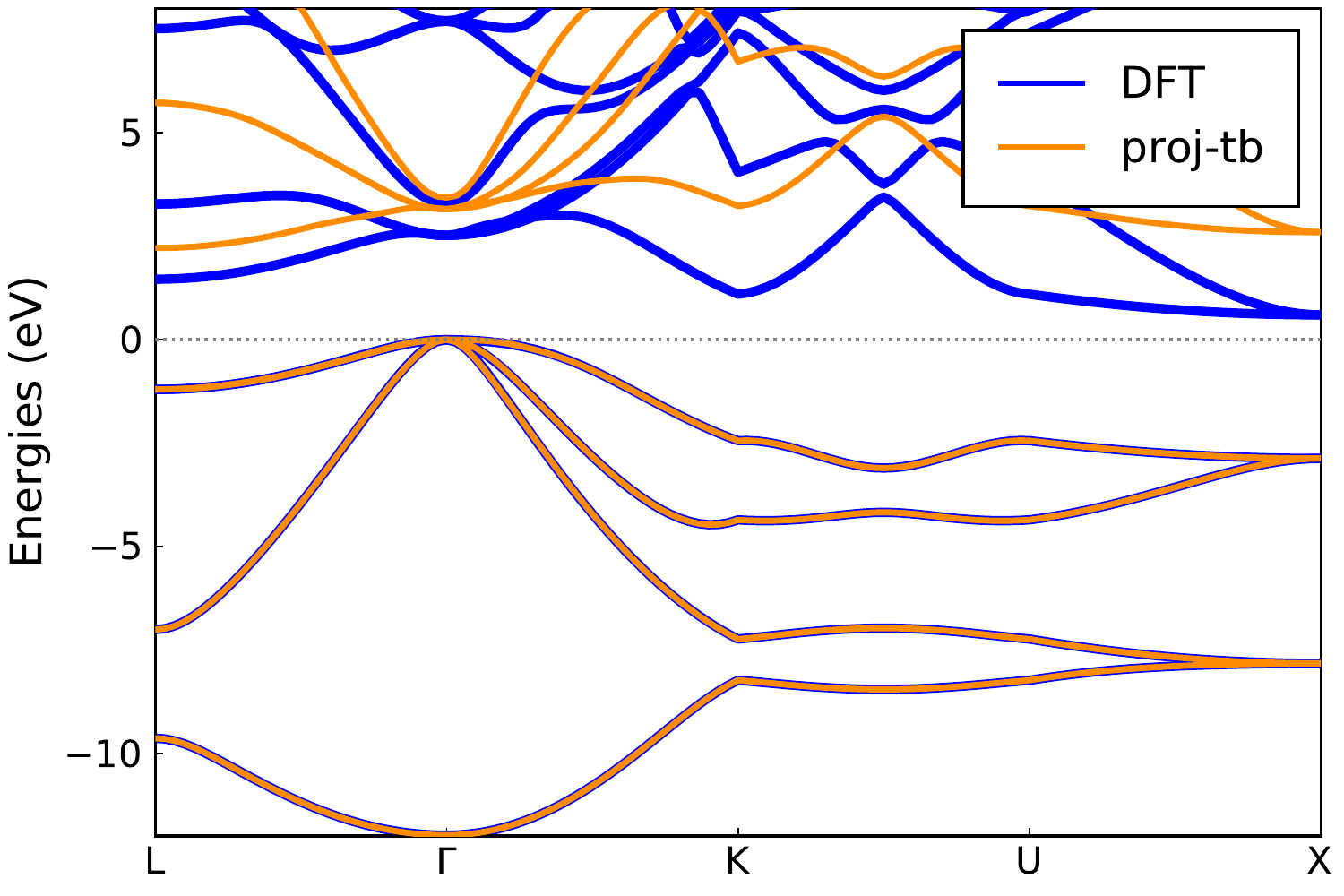}
\caption{\label{fig:si_dia} Band structure comparison between DFT (blue), and atomic projected tight-binding (orange) for Si in the diamond structure. The zero of energy is the valence band maximum.}
\end{figure}

\subsection{Implementation Details}

The projection method described above picks out the highest projectibility Hamiltonian for the set of $M$ atomic orbitals, and can be used to separate both semicore states and high energy states from the valence and conduction bands we wish to describe. Ideally it also maintains the symmetry of the tight-binding Hamiltonian. However, we note that the original selection of $N$-bands at each $k$-point can be a subtle source of symmetry breaking, as the $N$-th and $N\!+\!1$-th band can be degenerate, and selecting only one of these at random introduces unwanted symmetry breaking. Therefore, we make sure to throw away the highest eigenvalues at each $k$-point before projection.

A second problem can occur if the desired set of atomic orbitals includes high energy states that are not well described by the $N$ Kohn-Sham bands in the original DFT calculation. In this case, the trace of the $M \times M$ matrix $B B^\dag$ will be much less than $M$. This situation can be monitored and can usually be solved by increasing $N$.

A more serious difficulty is that the projection scheme does not reproduce even the occupied states exactly. While it is impossible to reproduce the larger set of unoccupied bands with only tight-binding orbitals, it is desirable to reproduce the occupied bands, and possibly the lowest conduction bands, for our eventual fitting procedure. Fortunately, the occupied orbitals are almost always well-described by atomic orbitals and our atomic-projected Hamiltonians require only small adjustments. 

We perform this adjustment by first we deciding on an energy range below which the eigenvalues should be exact by defining a smooth cutoff function $f(E)$ that is one below some cutoff energy and that goes to zero at higher energies. Then, we can adjust the TB eigenvalues to match the DFT eigenvalues while keeping the TB eigenvectors unchanged:
\begin{eqnarray}
H^{TB} = \Psi E^{TB} \Psi^\dag\\
H^{Adj} = \Psi E^{Adj} \Psi^\dag
\end{eqnarray}
Here, $\Psi$ are the eigenvectors, and $E^{TB}$ and $E^{Adj}$ are diagonal matrices of tight-binding and adjusted eigenvalues. The adjusted eigenvalues are \begin{eqnarray}
\epsilon_n^{Adj} = f(\epsilon_n^{DFT}) \epsilon_n^{DFT} + (1 - f(\epsilon_n^{DFT})) \epsilon_n^{TB},
\end{eqnarray}
where $\epsilon_n^{Adj}$, $\epsilon_n^{TB}$, and $\epsilon_n^{DFT}$ are the adjusted, tight-binding, and DFT eigenvalues, respectively. For this procedure to work, it is necessary to identify which DFT eigenvalue should be matched with each TB eigenvalue. We do this by comparing the energies and the eigenvector projections on the DFT bands to find the best match. We take the cutoff energy to be the lowest eigenvalue above the Fermi 
level, and the cutoff range is 3 eV.

In Fig.~\ref{fig:si_dia}, we show a comparison between the DFT eigenvalues for silicon in the diamond structure and our atomic projected tight-binding model, using the method described in this section. We can see that there is excellent agreement for the occupied eigenvalues, even for k-points along high symmetry lines but not in our original grid. However, there is much worse agreement for the conduction bands, with the tight-binding bands only tracing the general shape of the conduction bands. This is because there is significant mixing between these states and various unoccupied Si $s^*$ and $d$-states and other states that are not part of our model, which limits our ability to describe these states using solely atomic-like $s$ and $p$ orbitals. It may be possible to improve this agreement by including more orbitals, but this will increase the cost of the tight-binding calculations, undercutting the main motivation for using tight-binding in the first place. We leave models with more orbitals or other approaches to future work.

\section{\label{sec:fit}Fitting}

\begin{figure}
\includegraphics[width=3.4in]{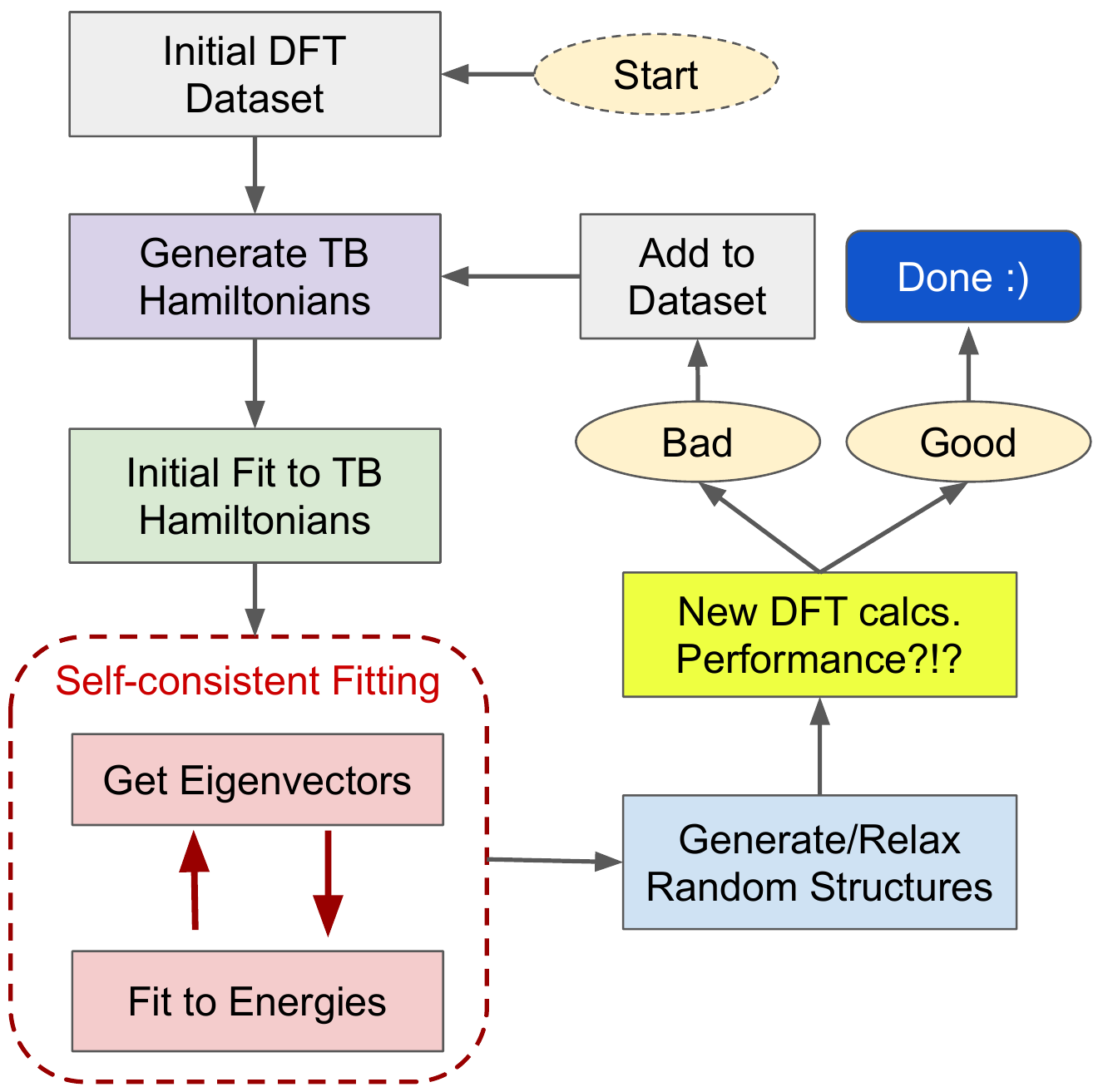}
\caption{\label{fig:fitting} Overview of the fitting process.}
\end{figure}

We fit tight-binding matrix elements to a set of DFT calculations by first doing a least squares fit to the set of initial DFT Hamiltonian matrix elements (see Sec.~\ref{sec:initial}). This is followed by another fit to the total energies and eigenvalues. A key part of our procedure is our recursive generation of new DFT fitting data to improve the model. We discuss these ideas in the following subsections. To orient the reader, an overview our procedure is presented in Fig.~\ref{fig:fitting}.

\subsection{Initial fitting}

Our initial fit is to the atomic-projected Hamiltonian matrix elements for a set of DFT calculations. Each DFT calculation contributes $n_k M^2$ matrix elements, where $n_k$ is the number of symmetry-reduced k-points and $M$ is the number of orbitals. The number of independent matrix elements is reduced by the Hermitian symmetry and any crystal symmetries. These matrix elements are arranged into a long vector of length $N_{TB}$. The charge self-consistency contributions (Sec.~\ref{sec:scc}) are subtracted from the matrix elements.

The set of descriptors is a $N_{TB} \times n_{param}$ matrix, where $n_{param}$ are the number of tight-binding model parameters that are relevant to the DFT calculations. These parameters include two-body terms (Eq.~\ref{eq:lag}), three-body terms (Eq.~\ref{eq:threebody2}), and onsite terms (Eq.~\ref{eq:onsite2}-\ref{eq:onsite3}). The entries of this matrix come from Fourier-transforming the tight-binding model of Sec.~\ref{sec:tb} for each material.

As noted in Sec.~\ref{sec:twobody}, all of our fitting parameters are linearly related to the Hamiltonian. The initial set of coefficients then comes from a simple linear least-squares fit of the model coefficients to the Hamiltonian matrix elements. This fit is generally good enough to produce reasonable looking band structures, but the total energies are not very accurate. A major difficulty with the fitting of total energies is that the bandwidth of a given material can be a dozen eV, but the energy differences between chemically relevant structures are on the order of 0.1 eV/atom, making it necessary to include the total energy directly in the fitting instead of indirectly through the Hamiltonian. We discuss this further in the next section.

We also fit the overlap matrices with the same procedure, except the overlaps are purely two-body interactions. The overlaps are simple to fit, and are fixed for the rest of the fitting.

\subsection{Self-consistent fitting}

Starting from our initial fitting described above, we seek to improve the model by focusing more directly on the observables we care most about, namely, the total energies and the occupied eigenvalues. Unlike the Hamiltonian itself, which as discussed in Sec.~\ref{sec:initial} can be always be arbitrarily modified by a choice of unitary transformation or disentanglement procedure, the energies and occupied eigenvalues are well-defined observables. Unfortunately, unlike the Hamiltonian matrix elements, our model is not linearly related to the energy or eigenvalues, which appears to pose a major difficulty for the efficiency of the fitting.

In order to overcome this difficulty, we first note that the eigenvalues $\epsilon_{nk}$ can be linearly related to Hamiltonian if we already know the eigenvectors $\ket{\psi_{nk}}$:

\begin{equation}
    \epsilon_{nk} = \bra{\psi_{nk}}  H_{k}  \ket{\psi_{nk} }.
\end{equation}
Therefore, we adopt a procedure where we use our current set of parameters to generate and diagonalize the current Hamiltonians for each material in our dataset, and then we use the resulting eigenvectors to generate the new set of descriptors, using the eigenvalues as the target data rather than the Hamiltonian. By adopting this approach, we can fit the eigenvalues using linear fitting. The problem is that the eigenvectors of the old parameters will not generally match the eigenvectors of the new parameters. Therefore, this procedure must be repeated many times to reach consistency between the eigenvectors and eigenvalues. As usual for self-consistent equations, we find that mixing the previous and new coefficients results in a more stable approach to the solution. Armed with the eigenvalues and eigenvectors, the total energy (Eq. \ref{eq:toten}) of each material can also be incorporated into the fitting straightforwardly.

One final difficulty is that when including charge self-consistency as in Sec.~\ref{sec:scc}, each material must be self-consistently solved with the current set of coefficients as an inner loop within our overall self-consistent procedure for fitting the coefficients.

\subsection{Generation of DFT datasets}

The fitting procedure described above requires a dataset of DFT calculations to fit. First, we generate datasets for the elemental systems and fit the elemental coefficients. Each element is fit separately. Then, keeping the elemental coefficients fixed, we generate datasets of binary compounds and fit the binary coefficients. The flexibility of our model enables us to fit binary compounds without sacrificing our ability to describe elements. In each case, we generate an initial dataset and then supplement it using a simple learning strategy to generate relevant new low energy structures.  

To generate the elemental datasets, we begin by substituting each element into a series of common elemental structures or molecules with small unit cells, \textit{e.g.} \textit{fcc}, diamond, etc., as well as a dimer. All structures have eight or fewer atoms, with one or two atoms the most common. For each structure, we consider a series of three to five volumes within $\pm 10\%$ of the equilibrium volume, for a total of $\approx$100 structures. We fit an initial set of coefficents to this dataset. Unfortunately, it is impossible to ensure \textit{a priori} that any such dataset has sufficiently varied structures so that the resulting model both a) describes low energy structures accurately and b) has no unphysical low energy structures. We therefore adopt a recursive learning strategy to systematically improve the model (see Fig.~\ref{fig:fitting}). We use the current model to search for new low energy structures and add them to the dataset.

Specifically, for each element, we generate several new structures with random lattice vectors and random atomic positions, ensuring that no atoms overlap\cite{airss}. These new structures have two or three atoms per unit cell, and we relax them using the tight-binding model. For each of the new relaxed structures, we perform a new DFT calculation and compare the new DFT energy to the TB energy. If the total energy per atom differs by more than a tolerance of roughly 0.1 eV/atom, we add the new structures to the dataset and restart the fitting. We continue adding new structures in this way until the out-of-sample performance on these low energy structures improves.

The procedure for binary compounds is similar, except that we have to consider differing stoichiometry as well. We start our dataset with a few common structural prototypes at a range of stoichiometries (\text{e.g.} rocksalt, CaF$_2$). We add a few extra common structures at chemically relevant stoichiometries for that binary pair, as well as any matching structures from the JARVIS-DFT database \cite{choudhary2020joint,choudhary2017high} with small unit cells. Finally, we include a dimer at several bond lengths, for a total of $\approx$100 starting structures. We again employ recursive learning, generating two or three new random structures at the following compositions: 2/2, 1/2, 2/1, 1/3, and 3/1. These structures are relaxed with the model and then compared to new DFT calculations. The process is iterated until the out-of-sample energies improve. In many cases, certain stoichiometries we consider may not be chemically relevant in equilibrium, but we want the model to give reasonable results for as wide of a range of materials as possible. 

This entire process results in a large dataset of DFT structures. We make the DFT calculations available on the JARVIS-QETB website (\url{https://jarvis.nist.gov/jarvisqetb/}). Details of the dataset generation and recursive procedure, including the prototype crystal structures for the initial dataset generation, are available on the ThreeBodyTB.jl code webpage and documentation. The fitted datasets themselves are available at \url{https://github.com/usnistgov/ThreeBodyTB.jl/tree/master/dats/pbesol/v1.2}.

\subsection{First principles details}

Our first principles DFT calculations are performed using Quantum Espresso~\cite{qe} code using the PBEsol\cite{pbesol} functional, which predicts accurate lattice constants and elastic properties of solids\cite{testingdft}. We describe atomic regions using slightly modified GBRV pseudopotentials\cite{gbrv, ultrasoft} as distributed with the code. The modifications are which atomic orbitals are included in the pseudopotential files for the purposes of the atomic projections, as well as minor modification of the oxygen pseudopotential. We perform calculations using a 45 Ryd. ($\approx$610 eV) plane-wave cutoff energy. We use k-point grids with a linear density of at least 29 per \AA$^{-1}$ and Gaussian smearing with an energy of 0.01 Ryd. ($\approx$0.136 eV), which we also set as the defaults for our tight-binding code. We perform only non-spin-polarized calculations. We use the JARVIS-tools\cite{choudhary2020joint} package to generate surface and vacancy structures.

\section{\label{sec:results}Results}

\subsection{Pedagogical example\label{sec:simple}}

\begin{figure}
\includegraphics[width=3.0in]{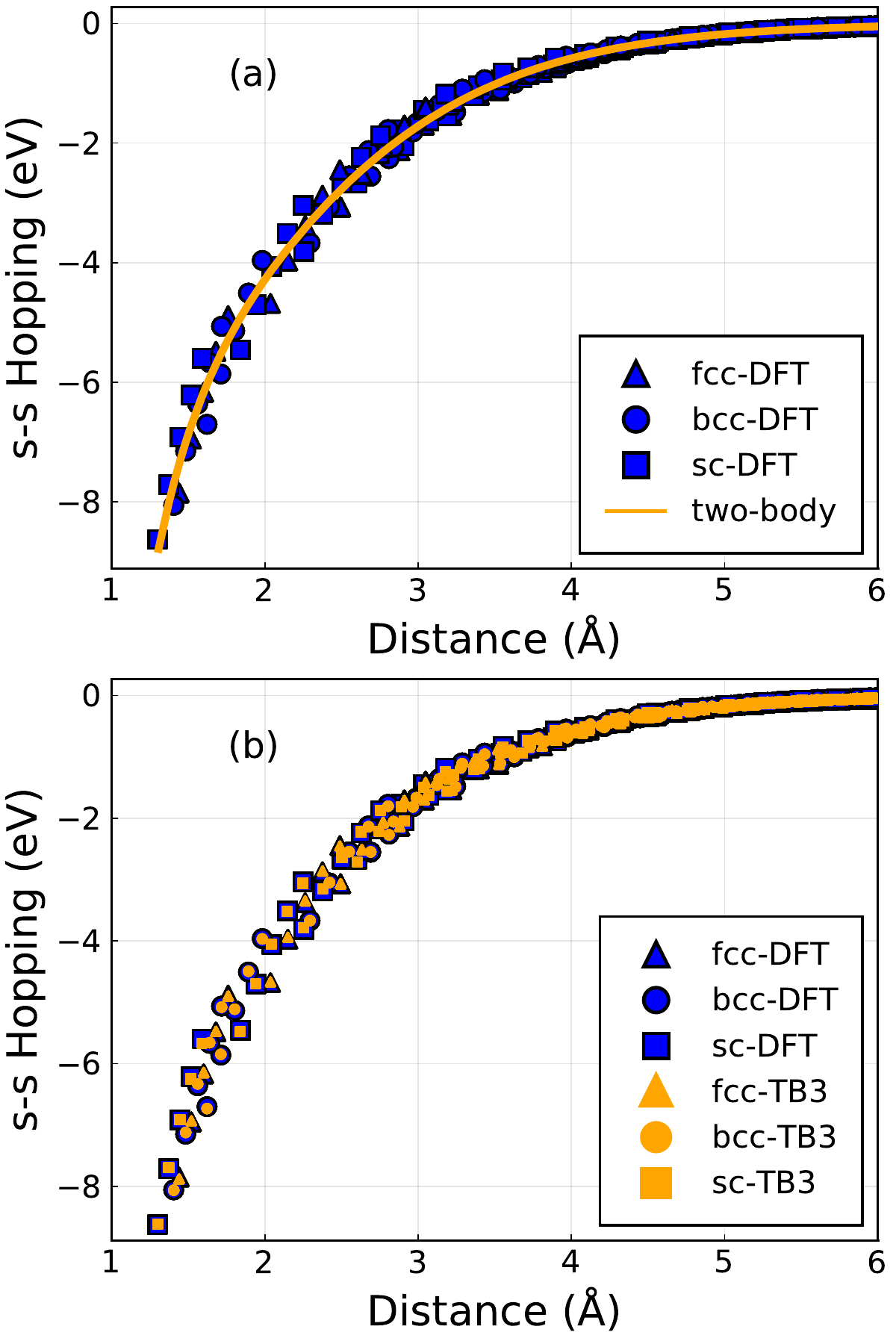}
\caption{\label{fig:example} Comparison of atom-projected DFT intersite $s$-$s$ Hamiltonian matrix elements for three hydrogen structures (blue symbols) with the a) two-body model, orange line, and b) three-body model (TB3), orange symbols. The three-body model points in b) are almost on top of the DFT results.  See text Sec.~\ref{sec:simple}.}
\end{figure}

We begin with a simplified pedagogical example that illustrates the power of the three-body tight-binding approach. For this example, we consider hydrogen atoms in three simple crystal structures, \textit{fcc}, \textit{bcc}, and \textit{sc} (face centered, body-centered, and simple cubic), at five volumes each. We describe hydrogen with a single isotropic \textit{s}-orbital, and for this example we fit directly to the atomic-projected Hamiltonian matrix elements per Sec.~\ref{sec:atomproj}. These Hamiltonian matrix elements are plotted as a function of distance in both panels of Fig.~\ref{fig:example} in blue symbols. We can see that there is strong decay with distance, but there is also a nearly 1.0 eV spread between the matrix elements of three different cubic structures at similar distances. Even within a single structure, the different shells of neighbors do not follow a single line versus distance.

If we fit a tight-binding model using purely two-body interactions as in Eqs.~\ref{eq:twobody1}-\ref{eq:lag}, the resulting intersite interactions between $s$ orbitals depend solely on distance. As shown in Fig.~\ref{fig:example}a, it is clearly not possible to describe all of these interactions accurately with purely two-body terms. However, by including three-body interactions as in Eqs.~\ref{eq:threebody1}-\ref{eq:threebody2}, the model can describe the additional variation in the matrix elements that comes from the differing local environments of the bonds. This can be seen in Fig.~\ref{fig:example}b, which shows almost perfect agreement between the three-body tight-binding model and the DFT matrix elements. This increase in flexibility and accuracy requires only three additional parameters in this case. 

\begin{figure}
\includegraphics[width=3.4in]{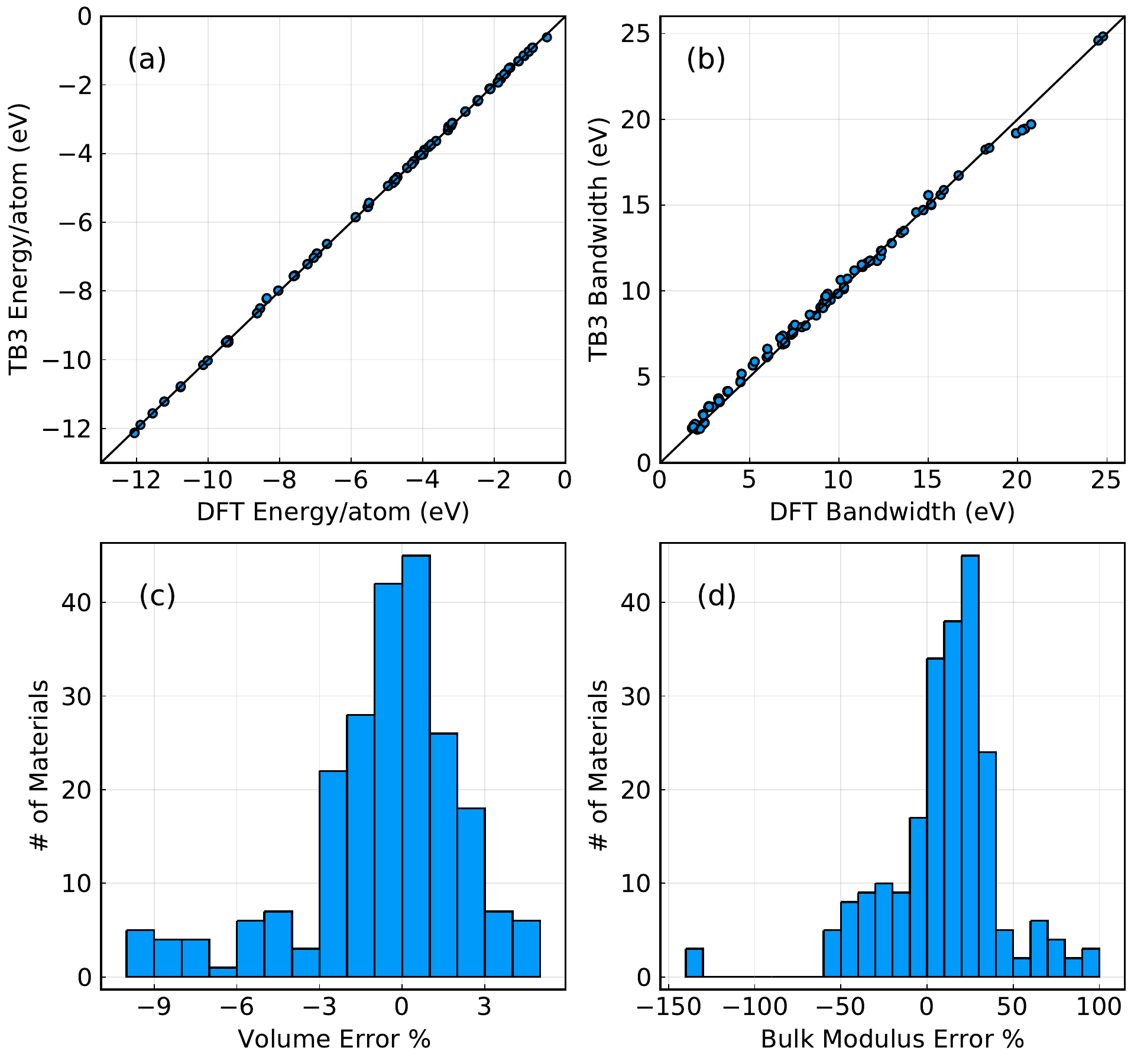}
\caption{\label{fig:el} Comparison of DFT and tight-binding properties for elemental systems: a) atomization energies (eV/atom), b) occupied electronic bandwidth (eV, see text), c) volume (absolute error percentage), d) bulk modulus (absolute error percentage).}
\end{figure}

\subsection{Bulk Structures}

We now present results demonstrating the accuracy of our model in reproducing and predicting bulk energies, volumes, bulk moduli, bandwidths, and band gaps. See supplementary materials Sec.~S4 for details on each structure we test. We separate our results into elemental systems, binary systems with small unit cells (2-6 atoms), and binary systems with large unit cells (9-10 atoms), only the last of which is an out-of-sample test. The structures we consider are the relevant bulk structures from the JARVIS-DFT database~\cite{choudhary2020joint}, which includes experimentally observed structures and other structures that are close to thermodynamic stability. We include a summary of these results in table~\ref{tab:summary}. The electronic bandwidth is defined as the difference between the valence band maximum and the lowest occupied states we include in our model. For ease of computation, the volume and bulk modulus are calculated for fixed internal atomic coordinates, \textit{i.e.} unrelaxed, and as in the entire paper, all calculations are non-spin-polarized.

We start by considering elemental structures. Because there are relatively few unique elemental structures that are observed experimentally, we do not have a separate test and training set for bulk elements (although see Sec.~\ref{sec:defect}). In Fig.~\ref{fig:el}, we present a comparison between the DFT and TB atomization energies, occupied state bandwidth, volume, and bulk modulus. The structures we consider are three-dimensional elemental solids. 

As can be seen in Fig.~\ref{fig:el}a, there is excellent agreement between the model and DFT atomization energies, which are a direct part of the fitting process. Fig.~\ref{fig:el}b shows that the TB model can also reproduce basic features of the band structure like the bandwidth. In Figs.~\ref{fig:el}c, we see that there is good agreement for the volumes, with most structures having less than 3\% error, which corresponds to only 1\% error in lattice constants. The bulk modulus, shown in Fig.~\ref{fig:el}d shows significantly more error. The bulk modulus is computed from six energy calculations between 94\% and 106\% of the equilibrium volume, and maintaining agreement with the first principles results over such a wide range is more challenging. In addition, some elemental structures include weak bonding between molecules, which is challenging for either our model or the underlying DFT to capture accurately.

\begin{figure}
\includegraphics[width=3.4in]{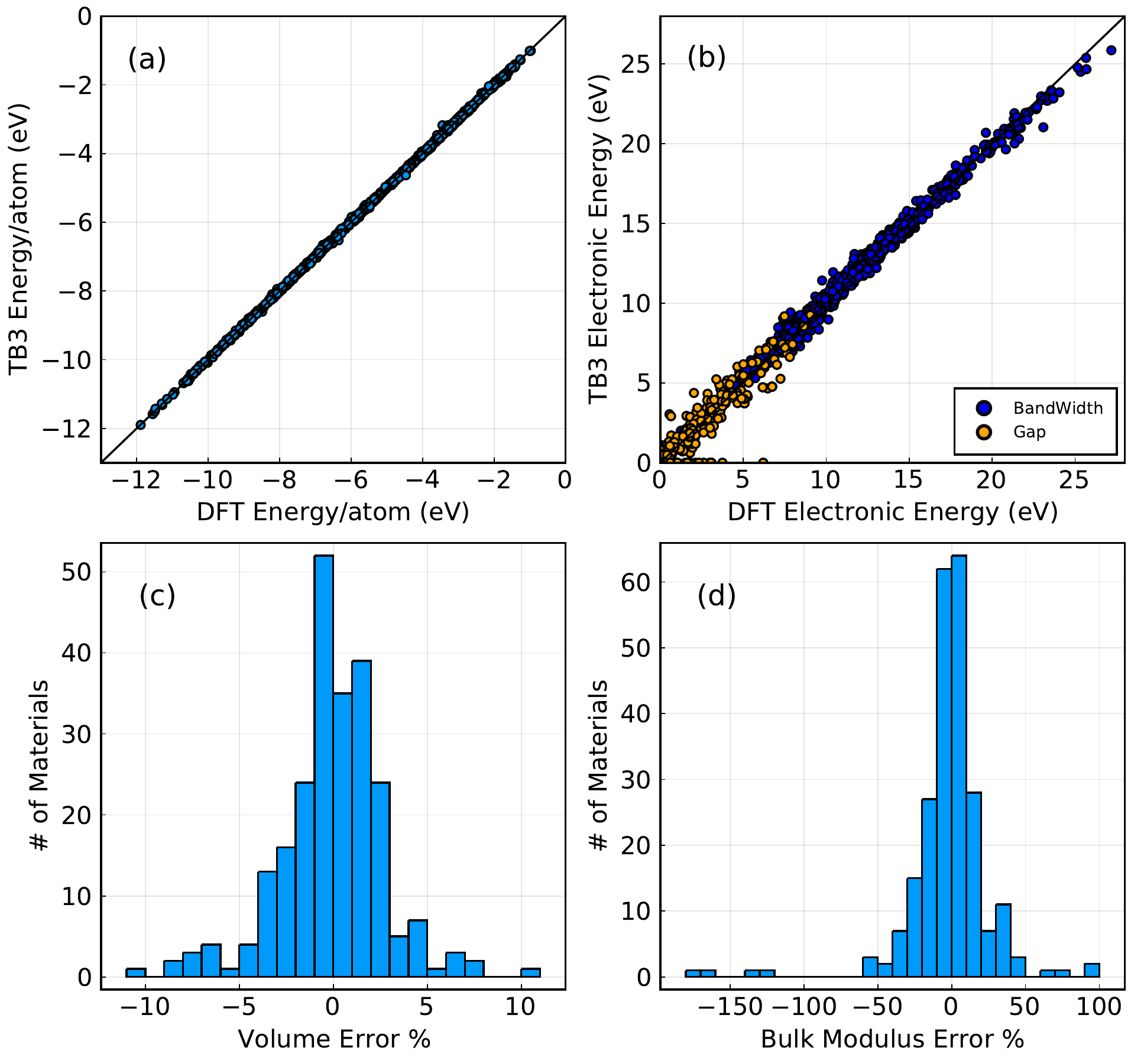}
\caption{\label{fig:bin26} Comparison of DFT and TB properties for in-sample binary compounds with two to six atoms per unit cell: a) atomization energies (eV/atom), b) occupied electronic bandwidth in blue and bandwidths in orange (eV, see text), c) volume (absolute error percentage), d) bulk modulus (absolute error percentage).}
\end{figure}

We move on to consider binary compounds. First, we consider binary compounds with two to six atoms per unit cell from the JARVIS-DFT database, which are again in-sample for our fitting procedure. The results, shown in Fig.~\ref{fig:bin26}, are again very promising, with excellent agreement for energies and bandwidths, good agreement for volumes, and reasonable agreement for the bulk modulus. In addition, in Fig.~\ref{fig:bin26}c, we show results for band gaps. Because our fitting procedure emphasizes the occupied eigenvalue and total energies, with a lower weight on unnoccupied bands, the band gaps are more challenging to fit quantitatively. Nevertheless, we find reasonable agreement between the DFT and TB band gaps.

\begin{figure}
\includegraphics[width=3.4in]{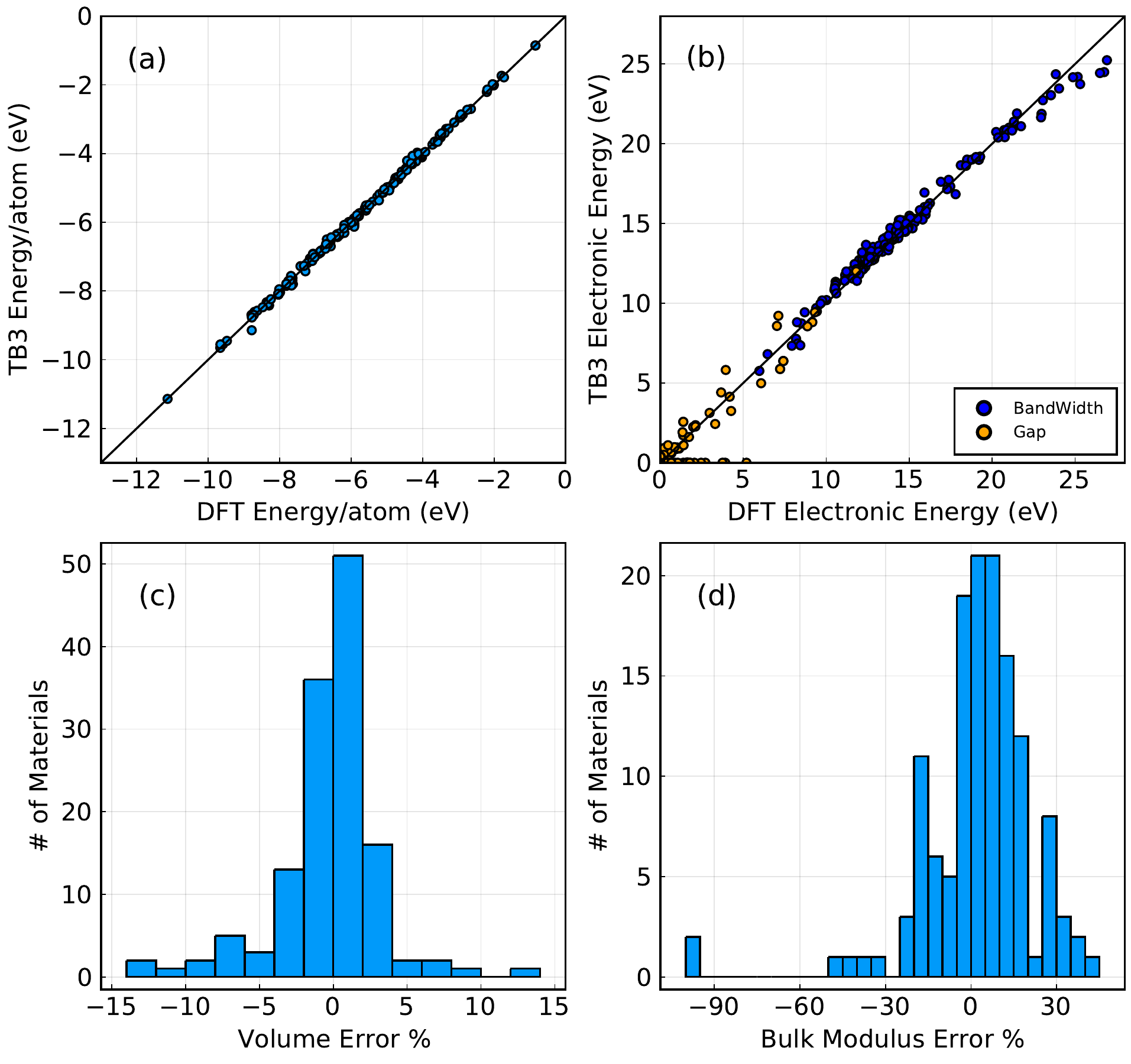}
\caption{\label{fig:bin910} Comparison of DFT and TB properties for out-of-sample binary compounds with nine to ten atoms per unit cell: a) atomization energies (eV/atom), b) occupied electronic bandwidth in blue and band gaps in orange (eV, see text), c) volume (absolute error percentage), d) bulk modulus (absolute error percentage).}
\end{figure}

Finally, we consider results for binary compounds with 9-10 atoms per unit cell from the JARVIS-DFT database, as shown in Fig.~\ref{fig:bin910}. None of these crystal structures are included in our fitting in any way, as we include only structures with eight or fewer atoms. Still, we find levels of agreement that are similar to our in-sample results. We find that the atomization energies (Fig.\ref{fig:bin910}a) are excellent, and the band gaps and bandwidths (Fig.~\ref{fig:bin910}b) are very good. The volume and bulk modulus errors (Fig.~\ref{fig:bin910}c-d) are also comparable to the in-sample data from Figs.~\ref{fig:el}-\ref{fig:bin26}. These results demonstrate the predictive power of our fit model over a wide range of chemistries, bonding types, and crystal structures.

\begin{table}
\caption{\label{tab:summary} Summary of model accuracy on bulk structures from the JARVIS-DFT database. Columns are absolute errors in atomization energy (eV/atom), volume (\% error), bulk modulus (\%), bandwidth (eV), and band gap (eV). Results are split into elements (in-sample) and small binary (2-6 atoms, in-sample) and large binary (9-10 atom, out-of-sample) unit cells.}
\begin{ruledtabular}
\begin{tabular}{lrrrrrr}
 & Energy & Volume & Bulk Mod. & Bandwidth & Gap \\ 
  &(eV/at.) & (\%) & (\%) & (eV) & (eV) \\
 \hline
Elements & 0.022 & 2.1 &  27 & 0.29 & $-$ \\
Binary & 0.018 & 2.0 & 17 & 0.30 & 0.46 \\
2-6 atoms & & & & & \\
Binary & 0.052 & 2.3 & 14 & 0.41 & 0.61 \\
9-10 atoms & & & & & \\
\end{tabular}
\end{ruledtabular}
\end{table}

\subsection{Band Structures}

\begin{figure}
\includegraphics[width=3.0in]{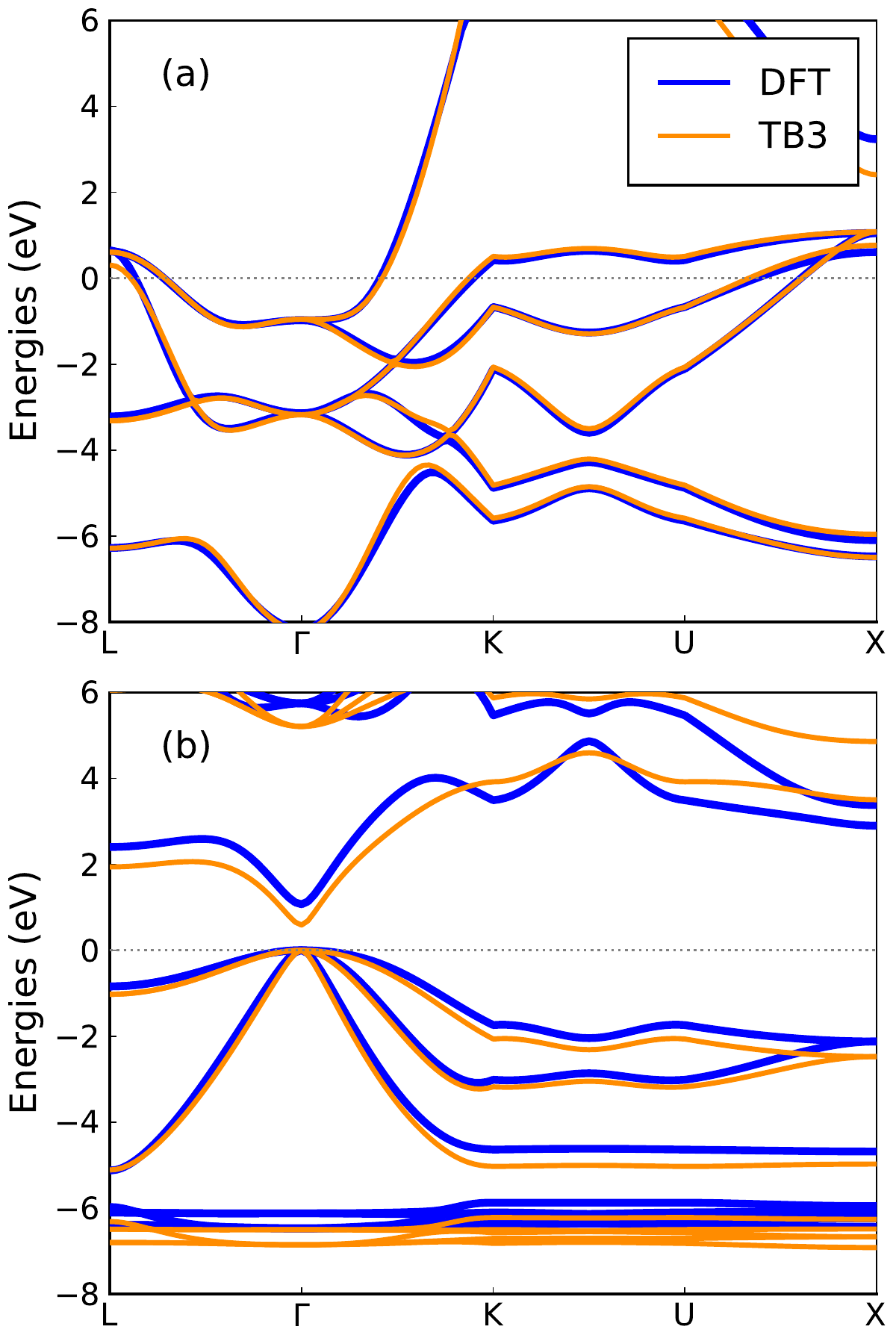}
\caption{\label{fig:bsins} In-sample band structure comparison between DFT (blue), and tight-binding (orange) for a) Rh in \textit{fcc} structure, and b) ZnSe in zinc-blende structure.}
\end{figure}

As discussed above, Figs.~\ref{fig:el}b-\ref{fig:bin910}b include statistical evidence of the accuracy of our model in reproducing electronic properties like the bandwidth and band gap. In this section, we present a few example comparisons between band structures calculated with tight-binding or directly with DFT. In Fig.~\ref{fig:bsins}, we show band structures for Rh in the \textit{fcc} structure as well as ZnSe in the zinc blende structure. These simple materials are both included in the relevant fitting datasets, and thus are in-sample predictions. As can be seen in the figure, we reproduce the occupied bands very well. The relatively localized $d$-states of Rh are very well described. The occupied Se \textit{p} states and lower energy Zn $d$-states again match the DFT band structure; although, the Zn $d$-states are shifted slightly. We also show reasonable agreement for the unoccupied bands, but the fit is less quantitatively accurate. 

\begin{figure}
\includegraphics[width=3.0in]{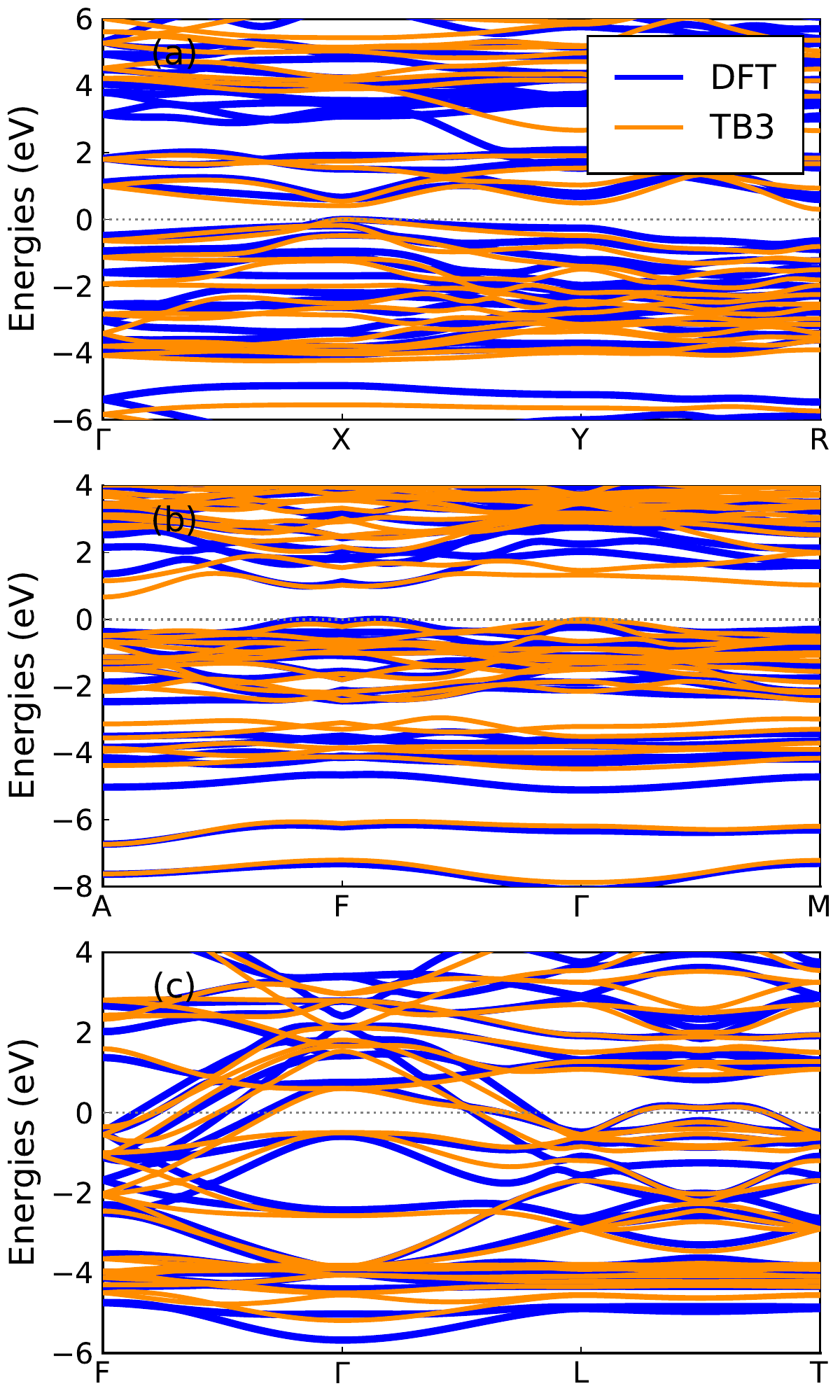}
\caption{\label{fig:bsoos} Out-of-sample band structure comparison between DFT (blue), and tight-binding (TB3, orange) for a) Ga$_4$Te$_6$, b) Ca$_5$P$_8$, and c) Au$_2$Bi$_8$ (see text).}
\end{figure}

In Fig.~\ref{fig:bsoos}, we show band structures for three materials with larger unit cells that are out-of-sample predictions: Ga$_4$Te$_6$ (\textit{Cc} space group, \textit{JVASP-22549}), Ca$_5$P$_8$ (\textit{C2/m}, \textit{JVASP-12962}), and Au$_2$Bi$_8$ (\textit{Fd-3m}, \textit{JVASP-101068}). Despite not being fit to these crystal structures, we are able to produce reasonable band structures in all three cases. Some of the bands are reproduced almost quantitatively, while others are shifted somewhat, but the averaged electronic properties are well reproduced with far less computational effort than full DFT calculations.

\begin{figure}
\includegraphics[width=3.4in]{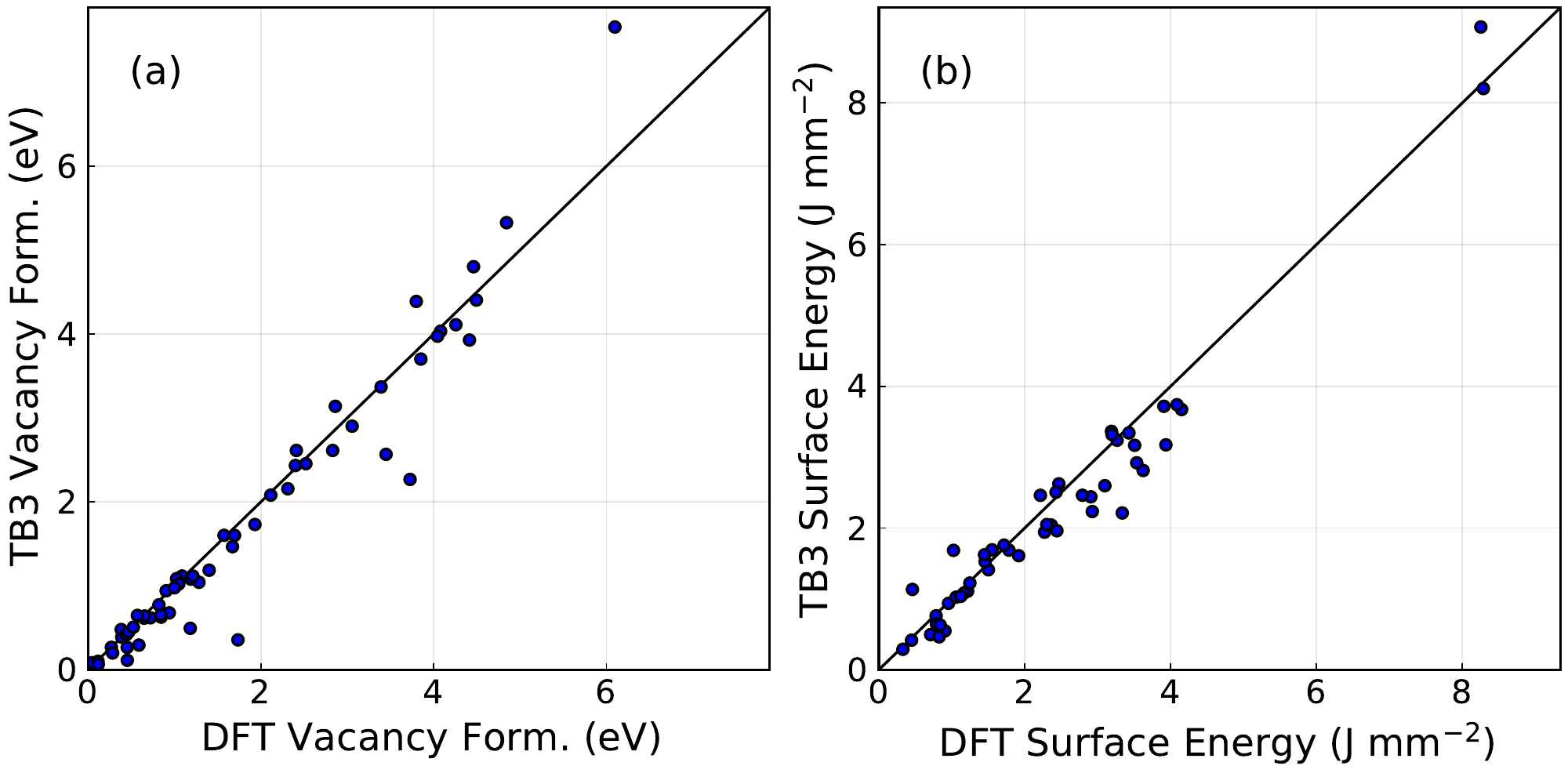}
\caption{\label{fig:defect_surf} Comparison of DFT and tight-binding calculations for unrelaxed a) point vacancy formation energy (eV) and b) $(111)$ surface energies ($J mm^{-2}$) of elemental solids.}
\end{figure}

\subsection{Defects and Surfaces\label{sec:defect}}

Thus far, we have only considered near-equilibrium properties of bulk materials. In this section, as a first step beyond these limitations, we consider vacancy formation energies and $(111)$ surface energies of elemental solids. For computational convenience, we only consider unrelaxed geometries. However, we also provide comparisons to calculated relaxed structures and experimental measurements in the supplementary materials when available\cite{medasani2015vacancy,popovic1974vacancy,haldar2017vacancy,pinto2006formation,freysoldt2014first,li2005defect,domain2005ab, kraftmakher1998equilibrium,ehrhart1991atomic,matter1979phase,chekhovskoi2012equilibrium,dannefaer1986monovacancy,schaefer1977vacancy,tzanetakis1976formation,satta1999first,gorecki1974vacancies,bourgoin1983experimental}, which show that relaxation effects are generally small in elemental systems. We note that none of the vacancy structures and none of the specific surfaces considered here are included in our fitting dataset, making these structures an out-of-sample test of the model. Our dataset does include thinner three to five atom slabs in the \textit{fcc} and \textit{bcc} structures.

We generate vacancy structures by first creating a supercell of the elemental ground state structure as necessary to ensure the defects are separated by at least 10\,{\AA}, and then deleting an atom. We calculate the vacancy formation energy as:
\begin{equation}
    V_{f} = E_{defect} - E_{ideal} + \mu.
\end{equation}
 where $E_{defect}$ and $E_{ideal}$ are the energies of the defect and ideal structures respectively, and $\mu$ is the chemical potential of the element in the same structure. A comparison between the DFT results and the tight-binding calculations are shown in Fig.~\ref{fig:defect_surf}a, which show good agreement in most cases across a wide range of defect energies.
 
 Next, we calculate the $(111)$ surface energies of the elemental solids in their respective reference structures and compare with DFT data in Fig.~\ref{fig:defect_surf}b. We generate surfaces with a 10\,{\AA} slab thickness and 15\,{\AA} vacuum padding during surface structure creation. We note that real surfaces can display significant reconstructions, but here we only consider ideal unrelaxed surfaces with a specific structure. We calculate surface energies as 
 \begin{equation}
    \gamma = (E_{surf} - \mu N_{at}) / (2 A),
\end{equation}
where $E_{surf}$ is the surface energy, $N_{at}$ is the number of atoms in the surface unit cell, $A$ is the surface area, and the factor of two is because slabs have two surfaces. As shown in Fig.~\ref{fig:defect_surf}b, we again find good agreement between the tight-binding results and the DFT surface energies. The raw data from the Fig.~\ref{fig:defect_surf} as well as a comparison to previous calculations and experiments is available in the supplementary information Sec.~S5.

\section{\label{sec:summary}Discussion and Summary}

The results of Sec.~\ref{sec:results} demonstrate that we are able to predict DFT energies and band structures using our parameterized tight-binding model including three-body interactions and self-consistent charges, with much reduced computation time (see Sec.~S1). This success shows our parsing of first principles electronic structures into at most three-atom effective interactions is a useful way to understand materials chemistry. In addition, we have indirectly demonstrated that the space of minimal atomic Hamiltonians is a smooth function of atomic positions even across a wide range of materials, which makes it possible to fit to our parameterized model in the first place. Also, because basic quantum mechanics and electrostatics are built directly into the formalism, we expect reasonable predictions when extrapolating beyond the training data. We note that the accuracy of our model in predicting the energies of bulk materials is comparable to state-of-the-art non-parametric machine learning models that do not directly include quantum mechanics\cite{vasudevan2019materials, ALIGNN, ml1, ml2, ml3, ml4}. It may be possible to improve predictions by combining the best features of both approaches, which has already been explored in a few studies\cite{mltb1, mltb2, mltb3}.

Still, our model has several shortcomings. First, for simplicity we currently include only non-spin polarized calculations, although there is no obvious problem with applying the approach to magnetic systems. Similarly, long-range London dispersion forces are missing from our underlying PBEsol DFT calculations, and thus not well described by our model, but not inherently problematic to the formalism. Second, there are remaining limitations of accuracy, especially in describing conduction bands or crystal structures that are very different from those in the training data. Finally, a more fundamental issue is that our use of three-body interactions means that applying our formalism to ternary (or quaternary, etc.) materials requires the inclusion of three-body terms between three different atom types. Such terms are not included in our current fitting set, which includes elemental and binary combinations only. We expect the importance of these terms to vary according to crystal structure, as we find that such three-body interactions are short-ranged. Adding ternary materials to our dataset systematically would require adding roughly an order-of-magnitude of DFT calculations to our already large dataset, but we may pursue a subset of materials.  

In summary, we have developed a tight-binding formalism that predicts the atomic-orbital Hamiltonian in terms of two-body and three-body interactions. The inclusion of three-body terms increases the model transferability and allows us to apply the same model to 65 elemental systems and any binary combination of those elements. We fit the model to a large dataset of DFT calculations, and we systematically generate new crystal structures until our model performs well on out-of-sample tests. To initialize the fitting process, we also develop a technique to generate an atom-projected tight-binding model for a single band structure. We demonstrate the effectiveness of this model in calculating total energies, volumes, elastic properties, and band structures of materials, as well as defects and surfaces. To enhance the utility and reproducibility of the current method, we provide software packages for the user to either directly use the current model parameterization for energy and band structure calculations, or to fit their own model. Finally, we have developed a publicly available database of the underlying DFT calculations.

%\begin{figure}
%\includegraphics[width=3.3in]{a.pdf}
%\caption{\label{fig:a} My caption.}
%\end{figure}

\bibliography{apssamp}% Produces the bibliography via BibTeX.

\pagebreak
\newpage
%\displaybreak
\clearpage
\onecolumngrid
\begin{center}
  \textbf{\large Supplementary Materials}\\[.2cm]
\end{center}
%\twocolumngrid

\setcounter{equation}{0}
\setcounter{figure}{0}
\setcounter{table}{0}
\setcounter{page}{1}
\renewcommand{\theequation}{S\arabic{equation}}
\renewcommand{\thefigure}{S\arabic{figure}}
\renewcommand{\bibnumfmt}[1]{[S#1]}
\renewcommand{\citenumfont}[1]{S#1}

% ****** Start of file apssamp.tex ******
%
%   This file is part of the APS files in the REVTeX 4.2 distribution.
%   Version 4.2a of REVTeX, December 2014
%
%   Copyright (c) 2014 The American Physical Society.
%
%   See the REVTeX 4 README file for restrictions and more information.
%
% TeX'ing this file requires that you have AMS-LaTeX 2.0 installed
% as well as the rest of the prerequisites for REVTeX 4.2
%
% See the REVTeX 4 README file
% It also requires running BibTeX. The commands are as follows:
%
%  1)  latex apssamp.tex
%  2)  bibtex apssamp
%  3)  latex apssamp.tex
%  4)  latex apssamp.tex
%
%\documentclass[amsmath,amssymb,prb,aps]{revtex4-2}

%\usepackage{graphicx}% Include figure files
%\usepackage{dcolumn}% Align table columns on decimal point
%\usepackage{bm}% bold math
%\usepackage{physics}
%\usepackage{amsmath}
%\usepackage{setspace}
%\newcommand{\del}{\,{\mathit{\Delta}}}
%\newcommand{\angstrom}{\mbox{\normalfont\AA}}
%\usepackage{hyperref}% add hypertext capabilities
%\usepackage[mathlines]{lineno}% Enable numbering of text and display math
%\linenumbers\relax % Commence numbering lines

%\usepackage[showframe,%Uncomment any one of the following lines to test 
%%scale=0.7, marginratio={1:1, 2:3}, ignoreall,% default settings
%%text={7in,10in},centering,
%%margin=1.5in,
%%total={6.5in,8.75in}, top=1.2in, left=0.9in, includefoot,
%%height=10in,a5paper,hmargin={3cm,0.8in},
%]{geometry}

%\begin{document}

\title{Supplementary materials for ThreeBodyTB}% Force line breaks with \\

%\author{Kevin F. Garrity}
% \email{kevin.garrity@nist.gov}
% \author{Kamal Choudhary}
% \affiliation{Materials Measurement Laboratory, National Institute of Standards and Technology, Gaithersburg MD, 20899}



%\begin{abstract}
Supplementary materials. Contains tables of data for various point vacancy and surface calculations and comparison with previous theory and experiment, as well as periodic table describing orbital choices.
%\end{abstract}

%\maketitle

%}

\section{\label{sec:level1}Timings}

In order for our scheme to be useful, it must be much faster to run a tight-binding calculation than an underlying DFT calculation. We expect the most computationally expensive step for large tight-binding calculations to be diagonalizing the Hamiltonian. This behavior is similar to DFT, but with a much smaller prefactor as we use many fewer basis functions (a minimal orbital basis vs plane-waves). When constructing the Hamiltonian, the three-body terms are more computationally expensive than the two-body terms; however, the added computational cost scales linearly with the number of atoms because the procedure is cutoff in real-space. Therefore the three-body terms do not add a large computational cost when studying large systems sizes.

\begin{figure}
\includegraphics[width=5.0in]{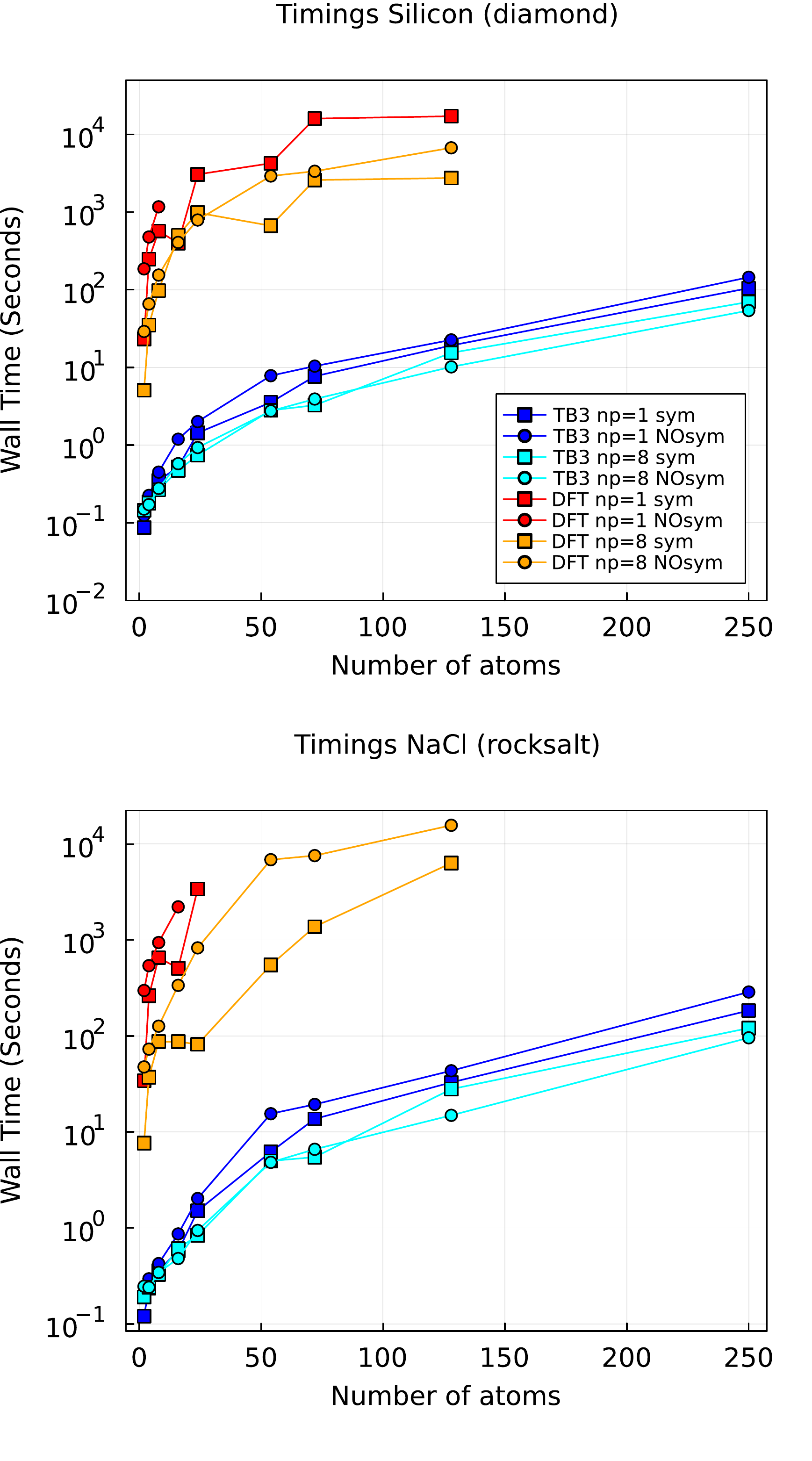}
\caption{\label{fig:timing} Timings in seconds for simple TB and DFT total energy calculations for Si (top) and NaCl (bottom) with varying numbers of atoms (note log scale on y-axis). Square symbols use symmetry, circles are slightly distorted structures with no symmetry. TB calculations with 1 processor (blue) or 8 processors (cyan) are up to 1000 times faster the equivalent DFT calculations with 1 (red) or 8 (orange) processors. }
\end{figure}

Exact coding timings depend on the hardware, compilation details, etc. Nevertheless, we provide example scaling relations in Fig.~\ref{fig:timing} for Si in the diamond structure and NaCl in the rocksalt structure, using our code and Quantum Espresso. We do calculations with and without taking into account symmetry to reduce the number of k-points, and with one and eight processors. Calculations were performed on Raritan cluster at NIST. We find that the tight-binding code is 500-1000 times faster than the DFT for these examples. Atoms with $d$-orbitals will have slightly slower results due to the higher number of orbitals per atom.

\section{\label{sec:level1}Comparison with DFTB+}

Density functional tight binding (DFTB, see discussion at \onlinecite{dftb_web}) and codes that implement it like DFTB+\cite{dftb, dftb_pt1,dftb_pt2,hourahine2020dftb+, dftbplus_web, porezag1995construction} are a longstanding method that is closely related to this work. In DFTB, a particular form for the Hamiltonian is derived by considering a second-order expansion of the Kohn-Sham energy with respect to charge density fluctuations. Typically, matrix elements for two-body interactions between actual orbitals are computed/interpolated, and repulsive terms between atoms are fit to the remaining energy. Usually at least charge self-consistency\cite{elstner1998self,elstner2007scc} is included like in this work, but many other terms have been included including spin-orbit, spin-polarization\cite{kohler2005density}, Van der Waals \cite{vdw}, etc. See discussions in Ref.~\onlinecite{hourahine2020dftb+}, for example. 

Because DFTB is a framework more than a specific Hamiltonian, set of datafiles, or procedures, it is difficult to compare in general with this work, but the methods are closely related. The major differences are 1) the procedure for generating (Slater-Koster) datasets from DFT calculations and 2) the number of datasets we make available. In this work, we include three-body (three-center) interactions between orbitals, which are rarely part of modern DFTB calculations (although see Ref.~\onlinecite{PhysRevB_40_3979, PhysRevB_56_6594,  PhysRevB_71_235101,elstner1998self, seifert2007tight, dftb,  porezag1995construction,fireball}). For the fitting, we fit all coefficients to DFT calculations rather than calculating and interpolating two-body Hamiltonian terms directly and only fitting the repulsive terms. Also, we simultaneously fit off-site and on-site interactions, rather than separately considering Hamiltonian and repulsive terms. 

The other difficulty in comparing the methods is that relatively few Slater-Koster parameter sets are available on the \textit{https://dftb.org/parameters} page. Those that are available were often compiled for specific use cases, often for organic molecules, and are not typically interoperable with parameter sets prepared for other purposes (see discussion on page). In this project, we instead seek to eventually create a universal and complete set of parameters for main group and transition metals appropriate for solid state materials, although we are currently limited to elemental and binary systems. Furthermore, we develop an automated testing procedure to test and improve our datasets, while the datasets available on \textit{DFTB.org} come from a variety of sources and methods.

\begin{figure}
\includegraphics[width=5.0in]{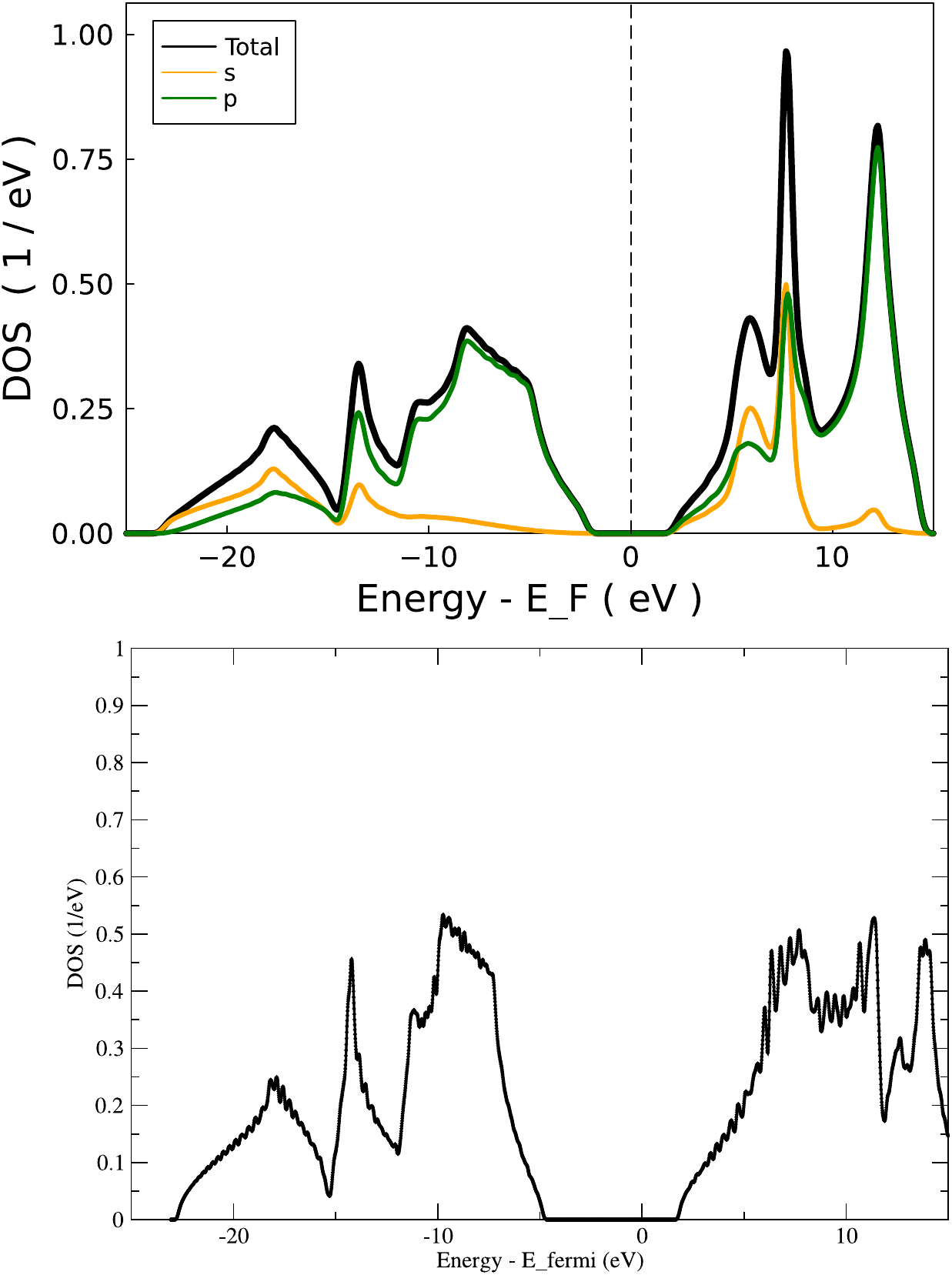}
\caption{\label{fig:dosboth} Top: diamond DOS performed with the ThreeBodyTB.jl code (this work). Bottom: same, performed with the DFTB+ code and the pbc-03 parameter set\cite{pbc03,RAULS1999459} }
\end{figure}

For cases where appropriate parameters are available for both this work and DFTB+, we expect reasonable agreement. As a simple example, Fig.~\ref{fig:dosboth} presents DOS plots for C in the diamond structure using the two codes. They show good agreement for the occupied orbitals.

\section{Two-body vs three-body model}

\begin{figure}
\includegraphics[width=5.0in]{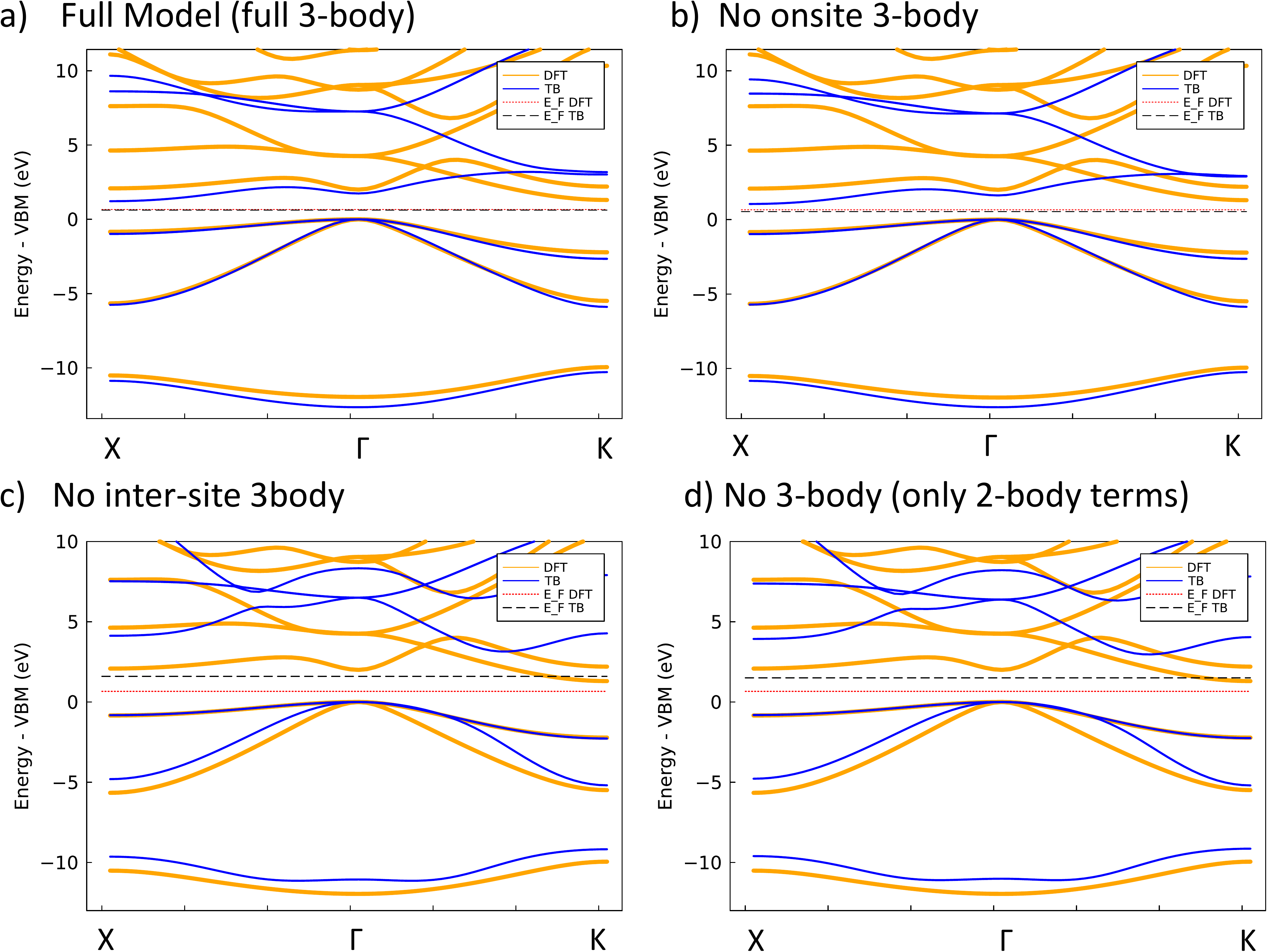}
\caption{\label{fig:2or3} AlAs band structure with a) full model b) no on-site three-body terms, c) no off-site three-body terms, and d) no three-body terms (only 2-body terms). Orange: DFT, Blue: TB. Energies aligned at valence band maximum. All panels include charge self-consistency.}
\end{figure}

As an example of typical magnitudes of three-body terms in the model, in Fig.~\ref{fig:2or3}, we compare the band structure of AlAs in the zinc blende structure with different terms in the model turned off, as compared to DFT. In panel Fig.~\ref{fig:2or3}a, we include the full model. Panel Fig.~\ref{fig:2or3}b turns off the onsite three-body interactions, Fig.~\ref{fig:2or3}c turns off the intersite three-body interactions, and Fig.~\ref{fig:2or3}d turns off all three-body interactions. We note that the effect of the intersite three-body interactions is much larger than the onsite interactions, which modify only fine details of the band structure. This observation of relative magnitudes is typical. The atomization energies are presented in table~\ref{tab:en} and show a similar pattern, although the three-body onsite term is quantitatively important. The relatively small magnitude of the onsite three-body term justifies of simplification of this term to be non-orbital dependent.

\begin{table}[hbt!]
\caption{\label{tab:en}Atomization energies of partial models (see Fig~\ref{fig:2or3})}
 \begin{tabular}{@{}lcc@{}}
 
\toprule
Method & Fig.~\ref{fig:2or3} panel & Atomization Energy (eV)\\
\hline
DFT & all & -9.68\\
Full & a & -9.68 \\
No onsite 3-body & b & -10.10\\
No intersite 3-body & c & -12.22\\
Only 2-body & d & -12.70\\
\botrule
\end{tabular}
\end{table}

\section{Crystal structure and detailed data}

Crystal structures and detailed data on each material used for testing is available at \url{https://doi.org/10.6084/m9.figshare.21158905.v1} and \url{https://doi.org/10.6084/m9.figshare.21158902.v1}. See also \url{https://jarvis.nist.gov/jarvisqetb/} for all the fitting data and \url{https://jarvis.nist.gov/jarvisdft/} for more details on each structure.

\section{\label{sec:level1}Data tables}

Various data tables related to Fig. 10 in the main text. The first table shows the data in the table, the second table is a comparison with some experimental and other theoretical literature values. We should expect only qualitative agreement with the literature values as we only consider unrelaxed unreconstructed structures and use the PBEsol\cite{pbesol} functional without spin-polarization.

\begin{table}[hbt!]
\footnotesize

\caption{Comparison of vacancy formation energies ($V$) in eV and (111) surface energies ($\gamma$) in Jm$^{-2}$ of solids with DFT and the TB model. All calculations are unrelaxed and non-magnetic, see details in main text. This is the data from Fig.~10. See also the following table. Not all of these surfaces are physically relevant, the main goal is to compare the model in an out-of-sample manner. See following table also.}
%\setstretch{0.7}
\begin{minipage}{174pt}
 \begin{tabular}{@{}llllllll@{}}
 
\toprule
Mat.&JVASP-ID& $V_{DFT}$ & $V_{TB}$ & Diff & $\gamma_{DFT}$&$\gamma_{TB}$ & Diff\\
\hline
Ag  &  14606  &  1.05  &  1.02  &  -0.03  &  1.17  &  1.08  &  -0.08  \\
Al  &  816  &  0.94  &  0.68  &  -0.26  &  0.91  &  0.55  &  -0.36 \\
As  &  14603  &  1.02  &  1.09  &  0.06   \\
Au  &  825  &  1.09  &  1.12  &  0.03  &  1.06  &  1.02  &  -0.03 \\
Ba  &  14604  &  0.47  &  0.45  &  -0.02  &  0.84  &  0.63  &  -0.21 \\
Be  &  834  &  2.83  &  2.61  &  -0.22  &  1.02  &  1.69  &  0.66 \\
Bi  &  837  &  0.65  &  0.64  &  -0.01  &   \\
Br  &  840  &  0.90  &  0.94  &  0.04  &   \\
C  &  25407  &  6.10  &  7.66  &  1.56  &  8.25  &  9.07  &  0.82 \\
Ca  &  25180  &  0.52  &  0.51  &  -0.02  &  1.25  &  1.22  &  -0.02 \\
Cd  &  14832  &  0.84  &  0.65  &  -0.19  &  0.79  &  0.76  &  -0.02 \\
Cl  &  25104  &  0.05  &  0.08  &  0.04  &   \\
Co  &  858  &  3.73  &  2.27  &  -1.46  &  2.44  &  1.96  &  -0.48 \\
Cr  &  861  &  4.26  &  4.11  &  -0.15  &  3.34  &  2.21  &  -1.12 \\
Cu  &  867  &  1.67  &  1.47  &  -0.20  &  1.45  &  1.62  &  0.17 \\
Fe  &  25142  &  4.41  &  3.93  &  -0.49  &  2.93  &  2.23  &  -0.69 \\
Ga  &  14622  &  1.19  &  1.08  &  -0.11  &  1.22  &  1.11  &  -0.11 \\
Ge  &  890  &  1.18  &  0.49  &  -0.69  &  2.47  &  2.62  &  0.16 \\
Hf  &  802  &  2.31  &  2.16  &  -0.16  &  2.91  &  2.44  &  -0.47 \\
Hg  &  25273  &  0.38  &  0.48  &  0.10  &  0.46  &  1.13  &  0.67 \\
I  &  895  &  0.65  &  0.62  &  -0.03  &   \\
In  &  898  &  0.59  &  0.29  &  -0.29  &  0.71  &  0.50  &  -0.21 \\
Ir  &  901  &  2.86  &  3.14  &  0.28  &  2.27  &  1.94  &  -0.33 \\
K  &  25114  &  0.12  &  0.06  &  -0.05  &  0.33  &  0.29  &  -0.04 \\
La  &  910  &  1.00  &  0.98  &  -0.02  &   \\
Li  &  25117  &  0.57  &  0.65  &  0.08  &   \\
Mg  &  919  &  0.85  &  0.63  &  -0.22  &  0.96  &  0.94  &  -0.02 \\
Mo  &  21195  &  3.85  &  3.70  &  -0.15  &  3.62  &  2.81  &  -0.81 \\
N  &  25250  &  4.08  &  4.03  &  -0.05  &  8.29  &  8.20  &  -0.09 \\
Na  &  931  &  0.27  &  0.27  &  0.00  &  0.45  &  0.42  &  -0.03 \\
Nb  &  934  &  3.06  &  2.90  &  -0.15  &  3.27  &  3.24  &  -0.03 \\
Ni  &  943  &  2.41  &  2.61  &  0.20  &  1.92  &  1.61  &  -0.31 \\
O  &  949  &  0.28  &  0.20  &  -0.08  &   \\
Os  &  14744  &  4.84  &  5.33  &  0.48  &  4.09  &  3.74  &  -0.35 \\
P  &  25144  &  1.21  &  1.12  &  -0.10  &   \\
Pb  &  961  &  0.45  &  0.27  &  -0.19  &  0.83  &  0.47  &  -0.36 \\
Pd  &  963  &  1.74  &  0.36  &  -1.38  &  1.78  &  1.69  &  -0.10 \\
Pt  &  972  &  1.93  &  1.73  &  -0.20  &  1.72  &  1.76  &  0.04 \\
Rb  &  25388  &  0.12  &  0.10  &  -0.01  &   \\
Re  &  981  &  4.46  &  4.80  &  0.34  &  3.91  &  3.72  &  -0.19 \\
Rh  &  984  &  2.52  &  2.46  &  -0.07  &  2.30  &  2.05  &  -0.25 \\
Ru  &  987  &  4.04  &  3.97  &  -0.07  &  3.54  &  2.92  &  -0.61 \\
Sb  &  993  &  0.82  &  0.77  &  -0.05  &   \\
Sc  &  996  &  1.57  &  1.60  &  0.03  &  2.43  &  2.51  &  0.08 \\
Se  &  7804  &  0.44  &  0.42  &  -0.03  &  1.46  &  1.53  &  0.07 \\
Si  &  1002  &  1.70  &  1.60  &  -0.09  &  3.19  &  3.36  &  0.17 \\
Sn  &  14601  &  0.72  &  0.62  &  -0.10  &  2.22  &  2.46  &  0.25 \\
Ta  &  1014  &  3.45  &  2.57  &  -0.88  &  3.51  &  3.17  &  -0.34 \\
Tc  &  1020  &  3.80  &  4.39  &  0.59  &  3.43  &  3.34  &  -0.09 \\
Te  &  25210  &  0.39  &  0.39  &  0.00  &  1.55  &  1.69  &  0.14 \\
Ti  &  1029  &  2.40  &  2.43  &  0.03  &  3.94  &  3.17  &  -0.76 \\
Tl  &  25337  &  0.45  &  0.11  &  -0.34  &  0.79  &  0.65  &  -0.14 \\
V  &  14837  &  3.39  &  3.37  &  -0.02  &  3.10  &  2.60  &  -0.50 \\
W  &  79561  &  4.49  &  4.40  &  -0.09  &  4.15  &  3.67  &  -0.48 \\
Y  &  1050  &  1.28  &  1.04  &  -0.24  &  2.36  &  2.04  &  -0.32 \\
Zn  &  1056  &  1.40  &  1.19  &  -0.22  &  1.12  &  1.04  &  -0.09 \\
Zr  &  14612  &  2.11  &  2.08  &  -0.03  &  2.79  &  2.46  &  -0.33 \\
\botrule
\end{tabular}
\end{minipage}
\end{table}

\begin{table}[hbt!]
\small
\caption{Comparison of vacancy formation energies (V) in eV and surface energies ($\gamma$) in Jm$^{-2}$ of solids with DFT \cite{medasani2015vacancy,popovic1974vacancy,haldar2017vacancy,pinto2006formation,freysoldt2014first,li2005defect,domain2005ab}, tight-binding model and experimental methods. \cite{kraftmakher1998equilibrium,ehrhart1991atomic,matter1979phase,chekhovskoi2012equilibrium,dannefaer1986monovacancy,schaefer1977vacancy,tzanetakis1976formation,satta1999first,gorecki1974vacancies,bourgoin1983experimental}. DFT and experimental values from literature, TB values from this work, unrelaxed and non-magnetic. The previous table is a more direct comparison between the TB model vs. DFT with the same functional in the same geometry. }\label{tab:vacancy}%
%\setstretch{0.7}
\begin{minipage}{174pt}
 \begin{tabular}{@{}llllllll@{}}
\toprule

Mat.&JV-ID& V$_{DFT}$                   & V$_{TB}$ &  V$_{Exp}$ &  $\gamma_{DFT}$ & $\gamma_{TB}$ & $\gamma_{Exp}$\\
Al &  816   &  0.77 \cite{medasani2015vacancy}  &  0.68  &  0.66 &  0.77 \cite{tran2016surface} &  0.55 & -\\
Ag &  14606  &  0.99 \cite{medasani2015vacancy}  &  1.02 &  1.06 & 0.76\cite{tran2016surface} & 1.08 & 1.32 \\
Au &  825   &  0.62 \cite{medasani2015vacancy}  & 1.12 & 1.02 & 0.71\cite{tran2016surface} & 1.02 & 1.54 \\
Cu &  867   &  1.27 \cite{medasani2015vacancy}  & 1.47 & 1.05 & 1.34\cite{tran2016surface} & 1.62 & 1.77 \\
Ni &  943   &  1.65 \cite{medasani2015vacancy}  & 2.61 & 1.4  & 1.92\cite{tran2016surface} & 1.61 & 2.01 \\
Pt &  972   &  0.96 \cite{medasani2015vacancy}  & 1.73 & 1.6 & 1.49\cite{tran2016surface} & 1.76 & 2.49 \\
Pd &  963   &  1.45 \cite{medasani2015vacancy}  & 0.36 & 1.85 & 1.36\cite{tran2016surface} & 1.69  & 2.0 \\
Rh &  984   &  1.99 \cite{medasani2015vacancy}  & 2.46 & 1.9 & 1.98\cite{tran2016surface} & 2.05 & 2.6 \\
Cr &  861   &  2.98 \cite{medasani2015vacancy}  & 4.11 & 2.0 & 3.44\cite{tran2016surface} & 2.21 & - \\
Mo &  21195   &  2.9 \cite{medasani2015vacancy}  & 3.7 & 3.6 & 2.96\cite{tran2016surface} & 2.81 & 2.9 \\
V &  14837   &  2.36 \cite{medasani2015vacancy}  & 3.37 & 2.07 & 2.70 \cite{tran2016surface} & 2.60 & 2.6? \\
W &  79561   & 3.54 \cite{medasani2015vacancy} & 4.40 & 4.0 & - & - & - \\
Co &  858   & 2.18 \cite{medasani2015vacancy} & 2.27 & 1.34 \\
Os &  14744   & 3.33 \cite{medasani2015vacancy} & 5.33 & 1.8 \\
Ti &  1029   & 2.15 \cite{medasani2015vacancy} & 2.43 & 1.55 \\
Tl &  25337   & 0.52 \cite{medasani2015vacancy} & 0.11 & 0.46 \\
Zn &  1056   & 0.5 \cite{medasani2015vacancy} & 1.19 & 0.54 \\
K &  25114   & 0.39 \cite{medasani2015vacancy}  & 0.06 & 0.34 \\
Na &  931  & 0.35 \cite{medasani2015vacancy}  & 0.27 & 0.26 \\
Si & 1002  & 3.6 \cite{medasani2015vacancy}  & 1.60 & 3.6 & 1.30 \cite{tran2016surface} & 3.36 & -\\
Fe & 25142 & 2.47 \cite{medasani2015vacancy} & 3.39 & 1.53\\
Mg & 919 & 0.81 \cite{medasani2015vacancy} &  0.63 & 0.79  & 0.76\cite{tran2016surface} & 0.94 & -\\
Ta & 1014 & 3.03 \cite{medasani2015vacancy} & 2.57 & 3.0  & 2.70\cite{tran2016surface} & 3.17 & 2.78\\
Rb & 25388 & 0.31 \cite{popovic1974vacancy} & 0.10 & 0.53\\
Pb & 961 & - & 0.27 & 0.58\\
C & 25407 & 7.6 \cite{li2005defect} & 7.66 & -\\
Ca &  25180  & 1.22 \cite{medasani2015vacancy}  & 0.51 & - & 0.46 & 0.48 & - \\
Ir & 901  & 1.87 \cite{medasani2015vacancy} & 3.14 & -\\
Nb & 934  & 2.99 \cite{medasani2015vacancy} & 2.90 & - & 2.34\cite{tran2016surface} & 3.24 & -\\
Tc & 1020 & 2.84 \cite{medasani2015vacancy} & 4.39 & -\\
Hf & 802 & 2.32 \cite{medasani2015vacancy} & 2.16 & -\\
Re & 981 & 3.68 \cite{medasani2015vacancy} & 4.80 & -\\
Ru & 987 & 3.0 \cite{medasani2015vacancy} & 3.97 & -\\
Sc & 996 & 1.95 \cite{medasani2015vacancy} & 1.60 & -\\
Ge & 890 & 2.62 \cite{pinto2006formation} & 0.49 & -\\
Zr & 14612 & 1.86 \cite{domain2005ab} & 2.08 & -\\
Y & 1050 & 1.95 & 1.04 & - & 1.11\cite{tran2016surface} & 2.04 & -\\
Cd & 14832 & - & - & - & 0.56\cite{tran2016surface} & 0.76 & -\\
Ga & 14622 & - & - & - & 0.51\cite{tran2016surface} & 1.11 & -\\
\botrule
\end{tabular}
\end{minipage}
\end{table}


\end{document}